 \documentclass[prl,amsmath,amssymb,twocolumn, showpacs, superscriptaddress,10pt]{revtex4-1}

\usepackage{amsmath}
\usepackage{hyperref}
\usepackage{graphicx}
\usepackage{amsfonts}
\usepackage{amsthm}
\usepackage{cases}
\usepackage{bm}
\usepackage{bbm}

\usepackage{color}
\definecolor{Blue}{rgb}{0.00, 0.00, 1.00}
\definecolor{Red}{rgb}{1.00, 0.00, 0.00}

\hypersetup{
    colorlinks=true,       
    linkcolor=red,          
    citecolor=blue,        
    filecolor=magenta,      
    urlcolor=cyan           
}

\newcommand{\nn}{\nonumber}
\newcommand{\be}{\begin{equation}}
\newcommand{\ee}{\end{equation}}
\newcommand{\bea}{\begin{eqnarray}}
\newcommand{\eea}{\end{eqnarray}}

\newcommand{\beq}{\begin{equation}}
\newcommand{\eeq}{\end{equation}}
\newcommand{\beqn}{\begin{eqnarray}}
\newcommand{\eeqn}{\end{eqnarray}}

\begin{document}

\title{Velocity and diffusion constant of an active particle in a one dimensional force field}

\author{Pierre Le Doussal}
\affiliation{CNRS-Laboratoire de Physique Th\'eorique de l'Ecole Normale Sup\'erieure, 24 rue Lhomond, 75231 Paris Cedex, France}
\author{Satya N. \surname{Majumdar}}
\affiliation{LPTMS, CNRS, Univ. Paris-Sud, Universit\'e Paris-Saclay, 91405 Orsay, France}
\author{Gr\'egory \surname{Schehr}}
\affiliation{LPTMS, CNRS, Univ. Paris-Sud, Universit\'e Paris-Saclay, 91405 Orsay, France}

\date{\today}

\begin{abstract}
We consider a run an tumble particle with two velocity states $\pm v_0$,
in an inhomogeneous force field $f(x)$ in one dimension.
We obtain exact formulae for its velocity $V_L$ and diffusion constant $D_L$ for arbitrary periodic $f(x)$
of period $L$. They involve the ``active potential'' which allows to define a global bias.
Upon varying parameters, such as an external force $F$, the dynamics undergoes transitions from non-ergodic trapped states, to various moving states, some with non analyticities in the $V_L$ versus $F$ curve.
A random landscape in the presence of a bias leads, for large $L$, to anomalous diffusion $x \sim t^\mu$, $\mu<1$, or to
a phase with a finite velocity that we calculate.
\end{abstract}



\maketitle

Persistent random walks, where a walker persists in the same direction for a finite time before changing direction, have
been studied extensively~\cite{Kac74, Lindenberg, Weiss_Lind,Weiss02,Masoliver_book}. The recent years have seen a resurgence of interest in this stochastic process
in a different reincarnation, namely the ``run and tumble particle'' (RTP), mostly in the context of active matter \cite{TC_2008,Fodor17,bechinger_active_2016,Sriram10,Cates15,cates_motility-induced_2015,Magistris15,SEB_16,SEB_17,Mallmin_18}. While several interesting
collective properties of interacting RTPs have been discovered recently, it was realised that even a single RTP exhibits 
rich and interesting static and dynamic behaviours \cite{Dhar_18,Sevilla18,ADP_2014,A2015,Cates_Nature,Malakar_2018,DM_2018,LDM_2019,GM_2019,EM_2018,M2019,kardar,MDMS20}. For example, the stationary state position distribution for an RTP 
in an external confining potential has been shown to deviate from the equilibrium Gibbs-Boltzmann form \cite{Dhar_18,Sevilla18}. Other interesting
questions such as the relaxation dynamics towards the stationary state in a confining potential \cite{Dhar_18}, the first-passage properties \cite{ADP_2014,A2015,Malakar_2018,DM_2018,LDM_2019,MDMS20,SK19} or 
the distribution of the current of non-interacting RTP's \cite{BMRS20} have been recently studied in the one-dimensional geometry. 
In this paper, we study a single RTP subjected to an external force periodic in space.
We show that, due to the presence of a finite persistence time, the position distribution under the periodic force exhibits a rich 
 and nontrivial behaviour, compared to the ordinary diffusion. In particular, we compute explicitly the velocity 
 and the diffusion constant of the RTP for an arbitrary periodic force $f(x)$.

The overdamped dynamics of the RTP is described by the stochastic evolution equation 
 \be
\frac{dx}{dt} = f(x) + v_0 \sigma(t)  \label{model}
\ee
where $f(x)$ is an external force and 
$\sigma(t) = \pm 1$ represents a telegraphic noise 
which switches from one state to another at a constant rate $\gamma$.

In free space on the line in the case where $f(x)=f$ is uniform, it is well
known that the dynamics of the active particle becomes diffusive at large time and can be effectively
described on large scale by a Langevin equation
\be \label{langevin}
\frac{dx}{dt} = f + \sqrt{2 D_0} \; \xi(t) 
\ee 
where the effective diffusion constant $D_0= v_0^2/(2 \gamma)$ and
the mean velocity is $V=f$. This effective description of \eqref{model} is valid above a 
characteristic persistence time $t^*=O(1/\gamma)$. In fact, the RTP dynamics
\eqref{model} converges to the Langevin dynamics \eqref{langevin}
in the limit where both $v_0 \to +\infty$, $\gamma \to +\infty$ 
with fixed $D_0$ \cite{footnote1}.

A natural question is what happens to this effective description 
when the RTP is subjected to an inhomogeneous force $f(x)$? In particular 
what is the mean velocity $V$ and the diffusion constant $D$ 
for arbitrary $f(x)$? In the case where $f(x)=- U'(x)$ with a
confining potential $U(x)$, there exists a stationary solution with 
zero current \cite{Horsthemke84,Klyatskin78a,Klyatskin78b,Lefever80,Hanggi95,Cates_Nature}. 
This stationary state was analysed in detail in \cite{Dhar_18}
for potentials of the type $U(x)=\alpha |x|^p$ and an interesting
"shape transition" in the stationary position distribution was found in the $(\alpha,p)$ plane. In that
case the RTP motion is bounded which corresponds to $V=0$
and $D=0$. In fact this stationary state is typically non-Boltzmann, 
which shows that the Langevin equation approximation breaks down.

In this paper we consider the RTP dynamics in Eq.~(\ref{model}) on an infinite line 
subjected to an arbitrary force landscape $f(x)$,
periodic in space, of period $L$, $f(x)=f(x+L)$ for all $x \in \mathbb{R}$. In this case, one would 
anticipate that, for small $f(x)$, there will not be any stationary position distribution and the particle
will keep on moving with time, with a non-zero speed $V_L$ and a non-zero diffusion constant $D_L$.   
One of the principal goals of this paper is to compute $V_L$ and $D_L$. But before we do that for the RTP, it
is useful to recall what happens for a simple diffusive particle (\ref{langevin}) subjected to this periodic force, which has been studied extensively~\cite{DerridaPomeau,DerridaLong,PLDV1995,GB1998}. 
In this case, the position distribution $P(x,t)$ satisfies the Fokker-Planck equation 
\begin{eqnarray}\label{FP}
\partial_t P = -\partial_x J \;, \;\; {\rm where} \;\;\; J = -D \partial_x P + f(x) P \;,
\end{eqnarray}
which, for bounded potential, does not have a normalisable steady-state solution. However, its periodised~version,  
\be
\tilde P(x,t)= \sum_{n=-\infty}^{+\infty} P(x+ n L,t) \;,
\ee
which satisfies the same Fokker-Planck equation (\ref{FP}),  
is known to reach a stationary limit $\tilde P(x,t) \to \tilde P(x)$ as $t \to +\infty$~\cite{DerridaPomeau,DerridaLong,PLDV1995,GB1998}.
Indeed $\tilde P(x,t)$ corresponds to the position distribution
of a diffusive particle on a ring of size~$L$. This stationary periodised solution
$\tilde P(x)$ can be computed explicitly by setting $\partial_t \tilde P = 0$ in (\ref{FP}),
looking for a solution with a non-zero constant current $J$. The constant $J$ can be
determined from the normalisation condition $\int_0^L \tilde P(x)\,dx = 1$. Knowing $J$, one 
can then find the velocity~$V_L$ from the general identity~\cite{DerridaLong}
\be \label{label}
V_L = \lim_{t \to +\infty} \frac{d}{dt} \int dx \, x \, P(x,t)  = J L \;,
\ee 
where $P(x,t)$ is the non-periodised distribution. Similarly, the diffusion constant $D_L$, defined as
\bea \label{def_DL} 
D_L = \frac{1}{2} \lim_{t \to +\infty} \frac{d}{dt}\left( \overline{x(t)^2}  - \overline{x(t)}^2 \right)
\eea 
where $\overline{x^k(t)} = \int dx \, x^k \, P(x,t)$, was also computed explicitly~\cite{DerridaPomeau,DerridaLong,PLDV1995,GB1998}.  
In addition, if the potential $U(x)$ is itself periodic, $U(0) = U(L)$, the current vanishes, $J=0$, and the
periodised solution converges to $\tilde P(x) = e^{- U(x)/D_0}/Z$ for $x \in [0,L]$ where $Z$ is a normalisation constant. Thus the dimensionless 
quantity which measures the ``tilt'' of the 
potential landscape, 
\be\label{def_bias_diff}
G_L = \frac{U(0)-U(L)}{D_0}
\ee
can be interpreted as an effective measure of the global bias which determines the sign of the 
velocity $V_L$.

\begin{figure}[t]
\centering
\includegraphics[width=0.7\linewidth]{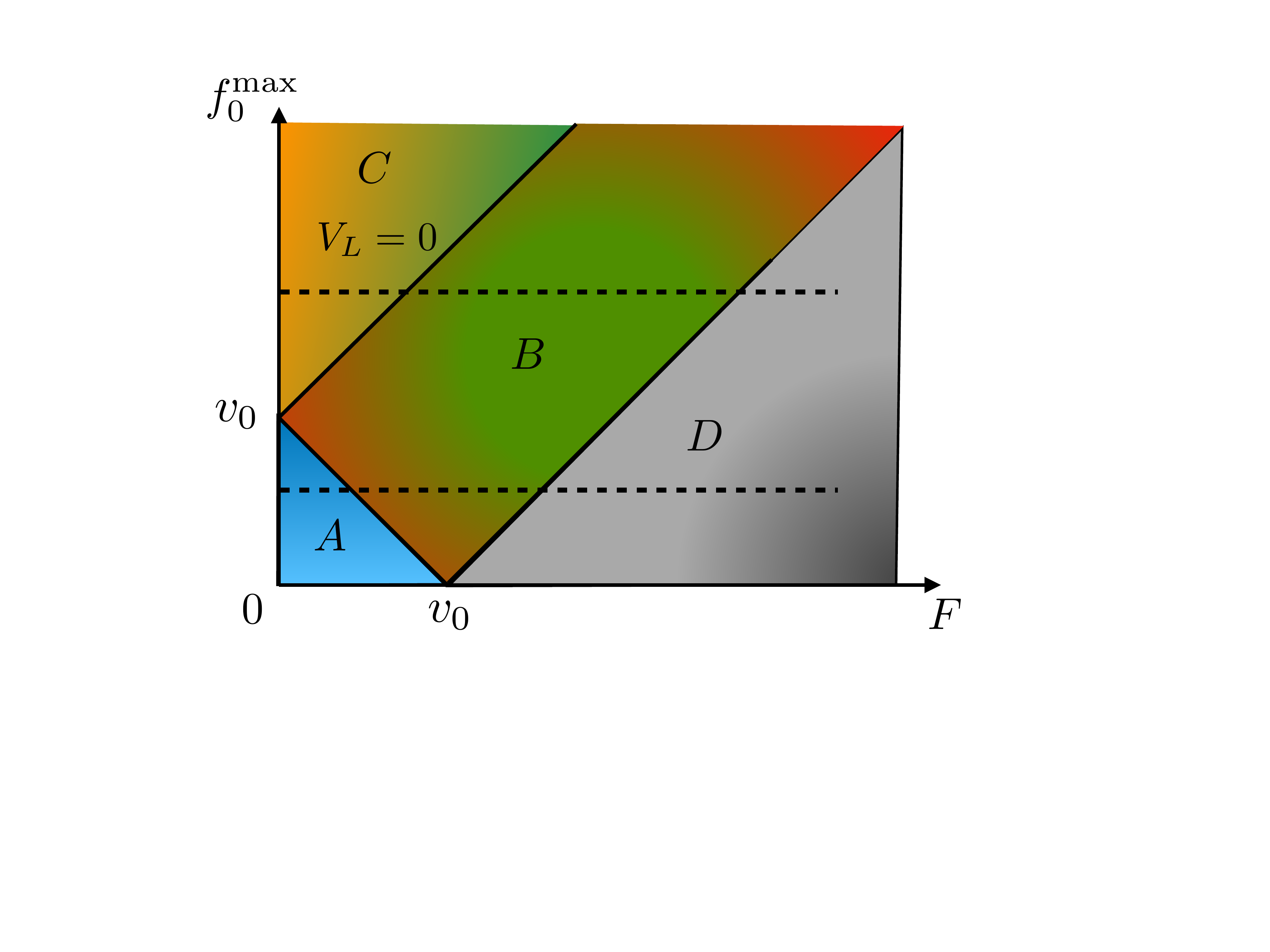}
\caption{Dynamical phase diagram of the RTP with $f(x)=f_0(x)+F$
as a function of the driving force $F$ and of the maximum
force $f_0^{\rm max}$ of the environment (assumed to
equal minus the minimum one). The $V_L$ versus $F$ characteristics
measured along the two dotted lines undergo different sequences
of transitions.}
\label{figABCD}
\end{figure}

In this paper, we carry out a similar procedure for the RTP (\ref{model}) subjected to this periodic force $f(x)=f(x+L)$ which we assume to be continuous.
However, due to the competition between the periodic force $f(x)$ and the noise (with a persistent memory) in Eq. (\ref{model}),
we show that one obtains a much richer behaviour for the periodised stationary solution leading to different phases and
transitions between them. Indeed we find four different phases (denoted by $A$, $B$, $C$ and$D$), depending on whether $f(x) = \pm v_0$ has real 
roots or not, leading to an interesting phase diagram shown in Fig. \ref{figABCD}. In addition, we also compute explicitly for any $L$,  
the stationary periodised solution $\tilde P(x)$, the velocity $V_L$ and the diffusion constant $D_L$. Furthermore, we also compute 
the mean first passage time to an arbitrary level $X$.

\begin{figure}[h]
\includegraphics[width=0.4\textwidth,angle=0]{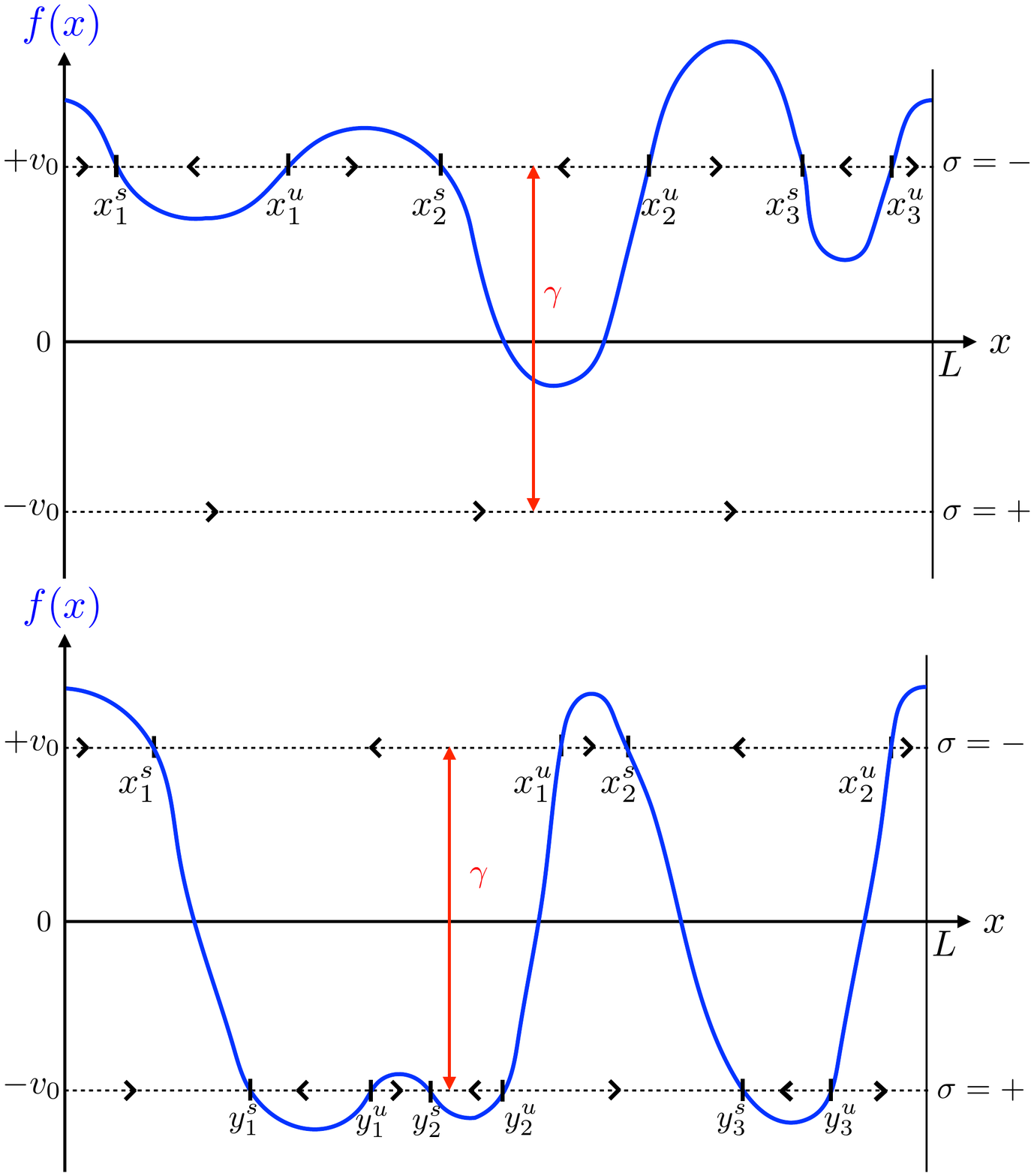}
\caption{Plots of $f(x)$ in a period $x \in [0,L]$ showing the stable $(s)$ and unstable $(u)$
fixed points, i.e. the roots of $f(x)=- \sigma v_0$, when the RTP is in the $\sigma=-1$ state (top line) 
and $\sigma=+1$ state (lowest line). The RTP moves along the arrows and
changes state with rate $\gamma$. {\it Top, phase $B$}: 
In the $-$ state the RTP moves left or right towards the fixed points $x_i^s$, and
in the $+$ state always to the right, leading to a mean velocity $V_L>0$. 
{\it Bottom, phase $C$:} the RTP ends up in either intervals $I_1=[x_1^s, y_1^s]$ or $I_2=[x_2^s, y^s_3]$, which are the supports
of the stationary measures (up to periodicity $L$), and $V_L=0$. Starting points in $[x_2^u-L,y_2^u]$ and $[x_1^u, y_3^u]$ 
end up in $I_1$ and $I_2$ respectively, with probability one. Starting 
in $[y_2^u, x_1^u]$ or $[y_3^u, x_2^u]$, the RTP ends up
randomly in either intervals}
\label{fig1}
\end{figure}

As mentioned above, the four phases are as follows (see also Figs. \ref{figABCD} and \ref{fig1}).

{\it Phase $A$}: $|f(x)|<v_0$ for all $x$. In this case the motion is unbounded and 
the stationary measure is smooth (if $f(x)$ is smooth).
We obtain a closed
formula for $V_L$ (see Eqs. (\ref{VL}) and (\ref{Psi})) and $D_L$ (see Eqs. (\ref{DL0}) and (\ref{DL})).

\begin{figure}[h]
\centering
\includegraphics[width=0.3\textwidth,angle=90]{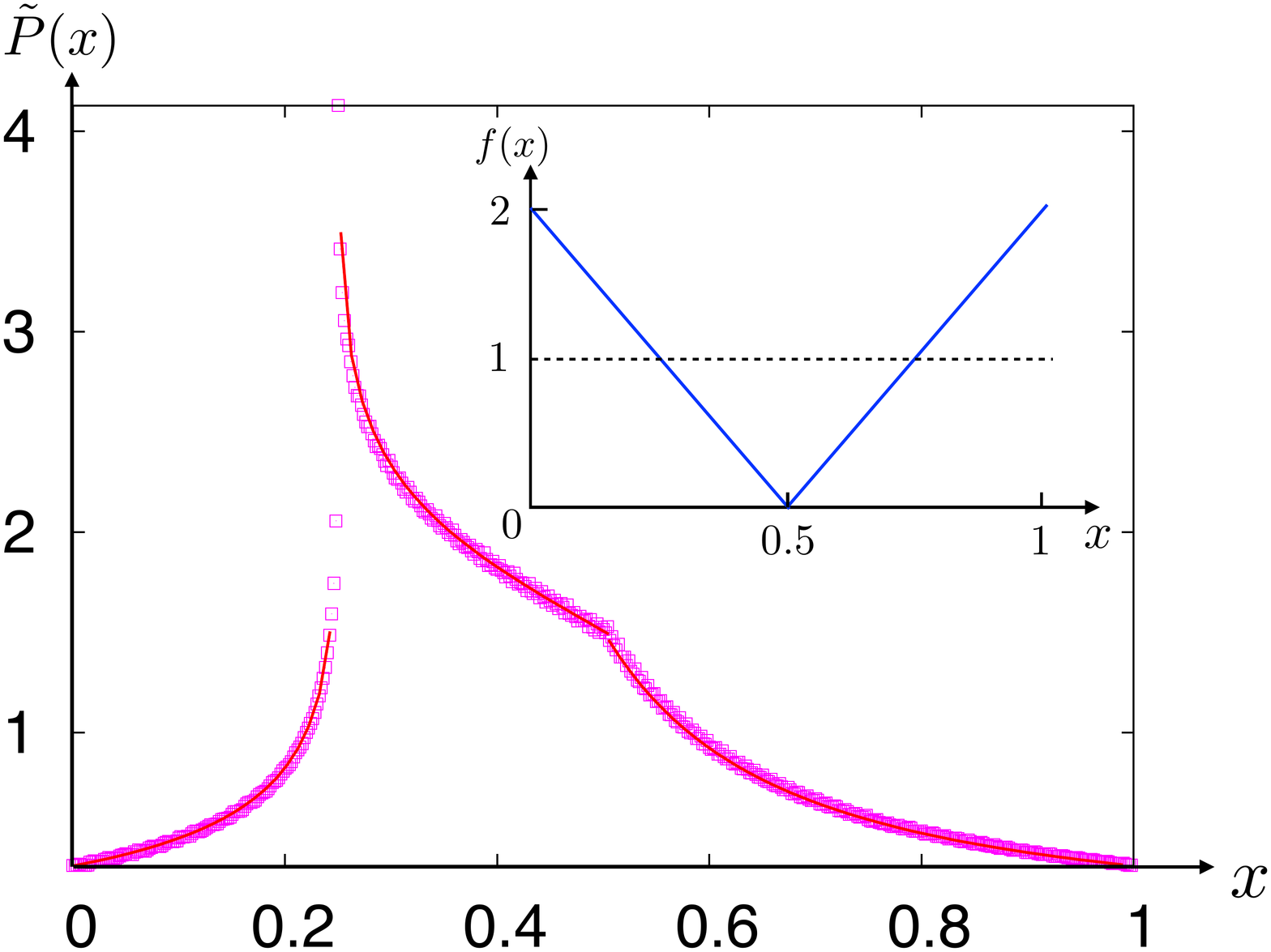}
\caption{Stationary distribution $\tilde P(x)$ versus $x$, for $f(x)=4 |x-\frac{1}{2}|$ (inset),
$\gamma=4$, $v_0=1$.
Squares: numerical simulation of the RTP (shown in period $[0,1]$). Solid line: analytical prediction
\eqref{plusieurs}. Note the logarithmic divergence at $x=1/4$. The angular
point at $x=1/2$ arises from the cusp in $f(x)$.}
\label{fig2}
\end{figure}

{\it Phase $B$}:  $f(x) > - v_0$ and there are roots $x_i$ (in increasing order) to $f(x)=v_0$ (see Fig. \ref{fig1} top panel). Out of these, 
every alternate ones (denoted by $x_i^s$ in Fig. \ref{fig1} top panel) are attractive fixed points for the RTP dynamics (\ref{model}) when it is
in the $-v_0$ state. The motion is a bit more complicated in this case. 
The position remains unbounded but the stationary periodised solution $\tilde P(x)$ 
has singular points (see Eq. (\ref{P2}) and also Fig. \ref{fig2}). We also obtain a formula for $V_L$ given in Eq. (\ref{VB}). 
A similar phase exists for the symmetric case where $f(x)<v_0$ and there are roots to $f(x) = -v_0$. 

{\it Phase $C$}: there exists roots to both $f(x)=+v_0$, denoted by $x_i$,
and to $f(x) = - v_0$, denoted by $y_i$ in increasing order (see Fig. \ref{fig1} bottom panel).
In this case the motion is bounded. The stationary periodised measure $\tilde P(x)$
has disjoint supports in a set of intervals with different weights depending
on the initial condition. The dynamics is non-ergodic in this case.

{\it Phase $D$}: $f(x)>v_0$ for all $x$. The RTP moves to the right 
in both $\pm v_0$ states. A similar situation arises for the symmetric
counterpart where $f(x)<-v_0$. 

Evidently, one can make transitions between these phases by tuning the maximum 
of the periodic force $f(x)$. One way to achieve this is to apply an additional
constant force $F$ on top of a periodic force landscape $f_0(x) = f_0(x+L)$. 
This amounts to setting $f(x)=f_0(x)+F$. Let $f_0^{\max}$ denote the maximum
of $f_0(x)$, for $x \in [0,L]$ and, for simplicity, we assume that $f_0^{\min} = -f_0^{\max}$. 
From our analysis, an interesting phase diagram emerges in the plane $(F,f_0^{\max})$ as shown in Fig.~\ref{figABCD}. 
The motion
undergoes transitions along the solid lines $f_0^{\rm max}+F=v_0$ ($A$ to $B$),
$-f_0^{\rm max}+F=v_0$ ($B$ to $D$) and $-f_0^{\rm max}+F=-v_0$ ($C$ to $B$).
As $F$ is increased along different lines (dotted lines in Fig. \ref{figABCD})
the velocity-force characteristics exhibits transitions, with $V_L=0$
in phase $C$, and non-analyticities in the $B$ phase as
new fixed points appear or disappear.

Another interesting question is whether, for the RTP, there exists a single global
measure $G_L$ of the bias as in the diffusive case in Eq. (\ref{def_bias_diff}). For this, it is
useful to define an "active external potential"
\be \label{Wdef} 
W(x)  = - 2 \gamma \int_{x_0}^x dy \frac{f(y)}{v_0^2-f^2(y)} \;,
\ee 
where $x_0$ is an arbitrary position. In the diffusive limit, $v_0, \gamma \to +\infty$ with fixed $D_0=\frac{v_0^2}{2 \gamma}$,
$W(x) \to U(x)/D_0$ converges to the standard external potential $U(x) = - \int_{x_0}^x f(x) dx$.  
We show that in phase $A$ the dimensionless global bias for the RTP, which determines the direction of the
velocity, can be expressed in terms of this active potential $W(x)$
\be \label{def_bias_RTP}
G_L = W(0) - W(L) =   2 \gamma \int_{0}^L dy \frac{f(y)}{v_0^2-f^2(y)}\;. 
\ee
Indeed we show that the sign of $V_L$ is the same as the 
sign of $G_L$ (and also $V_L$ vanishes when $G_L$ vanishes). 
Clearly, in the diffusive limit, Eq. (\ref{def_bias_RTP}) reduces to Eq.~(\ref{def_bias_diff}).
In addition, we show that in the small bias limit ($G_L\to 0$), 
the velocity satisfies an Einstein-like relation (within linear response in $G_L$)
%
\be \label{E} 
V_L \simeq D_L^{\rm zb} \, \frac{G_L}{L} \;,
\ee 
where $D_L^{\rm zb}$ denotes the diffusion constant $D_L$ 
in the case of
zero bias
[given below in \eqref{DL0}]. 
%


Let us first outline briefly our derivation of the main results. 
We first define $P_\pm(x,t)$ as the probability densities of the RTP to be in position $x$ at time
$t$ and in the state $\sigma(t)=\pm1$. They satisfy the pair of Fokker-Planck equations corresponding to Eq. (\ref{model})
\bea
&& \partial_t P_+ = - \partial_x [ (f(x) + v_0)  P_+]  -  \gamma P_+ + \gamma P_-  \label{P+evol.1}\\
&& \partial_t P_- = - \partial_x [ (f(x) - v_0)  P_- ] + \gamma P_+ -  \gamma P_-  \, . 
\label{P-evol.1}
\eea
The associated periodised distributions, $\tilde P_{\pm} (x,t) = \sum_{n} P_{\pm}(x+ n L,t)$,
satisfy the same pair of equations due to the periodicity of $f(x)$. We also define
the total probability $\tilde P(x,t)= \tilde P_+(x,t)+ \tilde P_-(x,t)$, as well
as the difference $\tilde Q(x,t)=\tilde P_+(x,t)-\tilde P_-(x,t)$, which then satisfy the
coupled Fokker-Planck equations
\bea \label{tildeP}
&& \partial_t \tilde P = - \partial_x J(x,t) = - \partial_x [ f(x) \tilde P + v_0 \tilde Q]  \;, \\
&& \partial_t \tilde Q = - \partial_x [ f(x) \tilde Q + v_0 \tilde P] - 2 \gamma \tilde Q  \;.\label{tildeQ}
\eea
At large time, assuming a stationary state to exist, we set $\partial_t \tilde P$ to zero in the first equation.
This implies that the probability current density $J(x,t)=f(x) \tilde P(x,t) + v_0 \tilde Q(x,t)$ converges to a constant 
$J= \lim_{t \to +\infty} J(x,t)$ independent of $x$. Hence, in the stationary state, we have $f(x) \tilde P + v_0 \tilde Q = J$ where
$J$ is yet to be determined. Eliminating $\tilde Q$ using this relation in Eq. (\ref{tildeQ}), and setting $\partial_t  \tilde Q = 0$, 
one obtains a first-order differential equation for~$\tilde P$
\be \label{eq_Ptilde} 
\frac{d}{dx} [(v_0^2 - f^2(x)) \tilde P(x) + J f(x)] 
+ 2 \gamma  J - 2 \gamma f(x) \tilde P(x)  = 0 \;.
\ee 
This equation can be explicitly solved for $\tilde P(x)$, using the periodicity condition $\tilde P(x+L) = \tilde P(x)$, see below.
Knowing $\tilde P(x)$ and $\tilde Q(x)$ from the relation $f(x) \tilde P + v_0 \tilde Q = J$, one gets the stationary distribution for each state~$\sigma~=~\pm$
\be \label{ppm_text} 
\tilde P_\pm(x) = \frac{\pm J + (v_0 \mp f(x)) \tilde P(x)}{2 v_0}  \;.
\ee 
Finally, the unknown constant $J$ is determined from the normalisation
condition $\int_0^L \tilde P(x) \, dx = 1$ and consequently the velocity $V_L = J\,L$ is obtained from Eq. (\ref{label}). 
The computation of the diffusion constant $D_L$ is a bit more cumbersome, but it can be derived from a generaliation
of the method used for the diffusive case \cite{DerridaPomeau,DerridaLong,PLDV1995}. The result for the zero-bias case
$G_L=0$ for phase $A$, is simpler and is given explicitly in Eq. (\ref{DL0}). The detailed derivation can be found in \cite{SM}.

%
{\it Phase $A$}. Let us first consider phase $A$, $|f(x)|<v_0$ for all $x$, in which the motion of the RTP is unbounded. 
Assuming a non-zero bias, i.e. $G(L) \neq 0$, and following the procedure outlined
above, we obtain the stationary distribution
\bea \label{P0} 
&& \tilde P(x) = \frac{2 \gamma J}{v_0^2 - f^2(x)} \Bigg(\int_0^L \frac{du \Phi_-(x,u)}{A_L} \nonumber
\\
&& ~~~~~~- \int_0^x du \Phi_-(x,u) - \frac{f(x)}{2 \gamma}\Bigg) \;,
\eea 
where we have defined 
\bea \label{Phipm} 
&& \Phi_{\pm}(x,u)=  \frac{v_0^2}{v_0^2 - f^2(u)} e^{\pm  (W(x)-W(u))} \;, \\
&& A_L = 1- e^{- (W(0)-W(L))} = 1 - e^{-G_L} \label{AL} \;.
\eea 
In the limit of zero bias $G_L \to 0$, one can show that $\tilde P(x) \to  \tilde A \, \Phi_+(0,x)$, for
$x \in [0,L]$ and $\tilde A$ is a normalisation constant.  
For arbitrary $G_L$, by determining $J$ from the normalisation condition $\int_0^L dx \tilde P(x)=1$, we get the velocity $V_L$ from
\eqref{label} 
\be \label{VL} 
\frac{1}{V_L}= \frac{1}{L} \int_{[0,L]^2} dx du \Psi(x,u) \left( \frac{1}{A_L}- \theta(x-u) \right) 
- \frac{G_L}{2 \gamma L} 
\ee
where we have further defined
\be \label{Psi} 
 \Psi(x,u)=  \frac{2 \gamma v_0^2 \, e^{- (W(x)-W(u))} }{(v_0^2 - f^2(x))(v_0^2 - f^2(u))} \;. 
\ee 
The formula \eqref{VL} for $V_L$ is exact for any~$L$ \cite{footnote11}.

To study the $L \to \infty$ limit, it is natural to assume that
$f(x)$ satisfies an ergodicity property, namely the existence of translational averages for local observables $O[f](x)$, denoted 
as $\langle O[f](x) \rangle_x = \lim_{L \to +\infty} \frac{1}{L} \int_0^L dx \, O[f](x)$. In addition [see Eq. (\ref{def_bias_RTP})] we assume that
\be \label{feff} 
\lim_{L \to +\infty} \frac{G_L}{2 \gamma L} =  \frac{f_{\rm eff}}{v_0^2} \;,
\ee 
where $f_{\rm eff} = \langle \frac{v_0^2 f(x)}{v_0^2 - f^2(x)} \rangle_x$ is an 
"effective active force" that arises from the global bias $G_L$.  
Without loss of generality, we assume $f_{\rm eff} >0$. Since $G_L \to  \infty$ from Eq. (\ref{feff}) it implies $\lim_{L \to +\infty} A_L=1$ from \eqref{AL}. Using
$A_L=1$, Eq. (\ref{VL}) can be re-arranged in a more compact form, leading to $\lim_{L \to +\infty} V_L=V$ where
\be \label{VelInfty} 
\frac{1}{V}= \int_0^{+\infty} dz \langle \Psi(x,x+z) \rangle_x - \frac{f_{\rm eff}}{v_0^2} \;.
\ee 
In addition, the diffusion constant $D_L^{\rm zb}$ for the case of zero bias $G(L)=0$, is obtained as
(see \cite{SM} for the general case)
\bea \label{DL0} 
\frac{D_0}{D_L^{\rm zb}} = \frac{1}{L^2} \int_0^L du \Phi_+(0,u) \int_0^L du' \Phi_-(0,u')  \;.
\eea 
In the large $L$ limit, $D^{\rm zb}_L \to D^{\rm zb}$ with
\bea \label{DL}
\frac{D_0}{D^{\rm zb}} = v_0^4 \left \langle \frac{e^{-W(x)}}{v_0^2 - f^2(x)} \right \rangle_x \times
\left\langle \frac{e^{W(x)}}{v_0^2 - f^2(x)} \right\rangle_x \;.
\eea 
This formula is valid provided each translational average
in \eqref{DL} converges. 
Finally, in the diffusive limit $v_0,\gamma \to +\infty$ 
with fixed $D_0= \frac{v_0^2}{2 \gamma}$, 
one can check that our formulae \eqref{VelInfty} and \eqref{DL} for $L \to \infty$ reduce to the diffusive 
results obtained in \cite{PLDV1995}.

{\it Phase $B$}. In this phase, there are $2 n$ roots to the equation $f(x)=v_0$ in a period $L$. 
Let us denote them by $x_i^s$ (stable) and $x_i^u$ (unstable), $i=1,\dots,n$,
with $f'(x_i^s) <0$ and $f'(x_i^u) >0$ (we assume for simplicity 
that $f(x)$ is differentiable). We choose the period such that
the roots are ordered as
$x^s_1 < x^u_1 <x^s_2 < \dots < x^u_n < x^s_{n+1}=x^s_1+L$, see Fig. \ref{fig1}.
The $x_i^{s,u}$ correspond respectively to
stable and unstable fixed points when the RTP is in the state $\sigma=-1$. 
The motion of the RTP in the state $\sigma=+1$ is always to the right.
Hence the RTP can not cross any of stable points to the left. Hence, we expect
a net drift with $V_L>0$ since the particle always spends a finite fraction of its time
in state $+$. The stationary measure can be computed from Eq. (\ref{eq_Ptilde}) and has
the form 
\bea \label{P2} 
\hspace*{-0.cm}&& \tilde P(x) = J \sum_{j=1}^n \mathbbm{1}_{[x^s_j, x^s_{j+1}]}(x) F_{x_j^u}(x) \\
\hspace*{-0.cm}&& F_y(x) = \frac{1}{v_0^2 - f^2(x)} 
\int^{y}_x du (2 \gamma + f'(u)) e^{2 \gamma \int_u^x dy \frac{f(y)}{v_0^2 - f^2(y)}} \nonumber \;.
\eea 
This expression is smooth around the unstable points $x_i^u$ but has
singularities near the stable points $x_i^s$ 
\be \label{phi} 
\tilde P(x) \sim (x^s_{i+1}-x)^{\phi_{i+1}} \quad , \quad \phi_{i+1} = -1 + \frac{\gamma}{|f'(x^s_{i+1})|} \;,
\ee
assuming that $f'(x)$ is continuous. From \eqref{ppm_text} we also obtain the
singularity associated to each state as
$\tilde P_{\pm}(x) \sim (x^s_{i+1}-x)^{\phi^\pm_{i+1}}$ 
with $\phi^-_{i+1} = \phi_{i+1}+1$ and $\phi^+_{i+1} = \phi_{i+1}$.
The velocity is then obtained from normalisation $\int_{x_1^s}^{x_1^s+L} dx \, \tilde P(x)=1$
and $J=V_L/L$ leading to the result for phase $B$
\be \label{VB} 
\frac{1}{V_L} = \frac{1}{L} \sum_{j=1}^n \int_{x^s_j}^{x^s_{j+1}} dx \, F_{x^u_j}(x)  \;.
\ee 

To illustrate phase $B$ we consider a simple example, $f(x)=4 |x-\frac{1}{2}|-1+ F$ with period $L=1$
(setting $v_0=1$). We first study the case $F=1$. The period is $[x_1^s, x_1^s+1]$ with $x_1^s=\frac{1}{4}$ 
and $x^1_u=\frac{3}{4}$ (see Fig. \ref{fig2}). We set $\gamma=4$, which 
leads to the simplest
expressions. By directly solving \eqref{tildeP}, \eqref{tildeQ} one finds that, due to the
angular points of $f(x)$ at $x=\frac{1}{2}$ and $x=1$, the solution 
$\tilde P=\tilde P_I$ takes different forms depending on the interval $I$ (see Fig. \ref{fig2})
\bea \label{plusieurs}
&& \tilde P_{[\frac{1}{4},\frac{1}{2}]}(x) = c_1 + \frac{J}{2} \log \frac{3-4 x}{4 x-1} ~,~ \tilde P_{[\frac{1}{2},1]}(x) = \frac{4 J x}{(1-4 x)^2} \nn \\
&& \tilde P_{[1,\frac{5}{4}]}(x) = c_2 + \frac{J}{2}  \log \frac{7-4 x}{5-4 x}  
\eea 
with $c_1= 2 J$, $c_2= \frac{J}{18} (8-9 \log 3)$ from continuity of $\tilde P$
at the angular points. Normalisation then leads to 
$V_L = J = \frac{18}{14+9 \log (3)}$. One can check these results agree with
a direct evaluation of the formula \eqref{P2} and \eqref{VB}
\cite{footnote2}. The result \eqref{plusieurs} is compared in Fig. \ref{fig2} with a numerical 
simulation of Eq. \eqref{model}. The exponent \eqref{phi} predicted for
the singularity at $x=x_1^s=\frac{1}{4}$ is $\phi=\frac{\gamma-4}{4}$,
in agreement with the logarithmic divergence in \eqref{plusieurs} for $\gamma=4$.
The case of general $\gamma$ is solved in \cite{SM}.

{\it Phase $C$}. In this phase, within the period $L$, there are $2 n$ solutions to $f(x)=v_0$, called $x^{s,u}_i$, $i=1,\dots,n$ ordered
as in phase $B$, and $2 m$ roots to $f(x)=-v_0$, called $y^{s,u}_j$, $j=1,\dots,m$ ordered
similarly with $f'(y_j^s) <0$ and $f'(y_j^u) >0$. The roots $y^{s,u}_j$ correspond respectively to
stable and unstable fixed points when the RTP is in the state $\sigma=+1$.
As a result the particle motion is now bounded, as one can check in the bottom panel of Fig.~\ref{fig1}. 
We now detail the structure of the possible stationary states.
Let us define ${\cal S}$ the subset of indices $i$ such that $[x_i^s, x_i^u]$ contains at least one
$y_j^s$, and define $k(i) = {\rm min}(j | \, x_i^s < y^s_j < x_i^u)$. The stationary distribution $\tilde P(x)$
has zero current $J=0$, and both the velocity $V_L$ and the diffusion constant $D_L$ are zero.
It reads
\bea \label{PC} 
&& \tilde P(x) =  \sum_{i \in {\cal S}}  A_i \frac{\mathbbm{1}_{[x^s_i, y^s_{k(i)}]}(x)}{v_0^2 - f^2(x)} 
 e^{2 \gamma \int_{u_i}^x dy \frac{f(y)}{v_0^2 - f^2(y)}} \;,
\eea 
where $u_i$ can be chosen as the midpoint $u_i=\frac{x^s_i+y^s_{k(i)}}{2}$. 
The stationary measure has a support made of a collection of disjoint intervals
and is zero elsewhere, which correspond to ``downwards travels'' of $f(x)$, as
represented in the bottom panel of Fig.~\ref{fig1}. The coefficients $A_i$, $i \in {\cal S}$, are however
determined by the initial condition, together with the normalisation condition.
Hence if there is more than one element in ${\cal S}$ the system is non-ergodic.

{\it Phase $D$.} In this case there are no roots to $f(x) = \pm v_0$. However there is a global
bias and the velocity $V_L \neq 0$. It turns out that both the stationary $\tilde P(x)$ and
$V_L$ are given by exactly the same formula as in phase $A$, namely by Eqs. (\ref{P0}) and (\ref{VL}) respectively.

In the limit $\gamma \to 0^+$, transitions being rare, the velocity simplifies (in all phases) as
$V_L \simeq \frac{1}{2} (V_+ + V_-)$, where $V_\sigma = L/\int_0^L \frac{dx}{f(x) + v_0 \sigma}$ is the velocity of an RTP frozen in state $\sigma=\pm 1$, with $V_\sigma=0$ if a root to 
$f(x)=- v_0 \sigma$ exists \cite{SM}.

{\it Transitions and velocity force characteristics.} As mentioned earlier, dynamical transitions
can occur between these phases as some external parameters are varied, such that $f(x)$ crosses
the levels $\pm v_0$, see e.g. Fig. \ref{figABCD}. Let us give a concrete example of this transition
for the model $f(x) = f_0(x) + F$ with $f_0(x)=4 |x- \frac{1}{2}|-1$ for $x \in [0,1]$ and we set $v_0=1$ as well as $L=1$. 
Clearly, in this case, $f_0^{\max} = v_0 = 1$. If we now vary $F$, we move along the horizontal line $f_0^{\max} = v_0$ 
in the phase diagram in Fig. \ref{figABCD}. For any $F>0$, the system is in phase $B$, a special case of this
was discussed before for $F=1$. However, exactly at $F=0$ the system is in phase $C$. Thus the critical
point in this example is exactly at $F=F_c=0$. As $F \to 0$, the velocity $V_L$   
vanishes as a power law $V_L \sim (F-F_c)^\beta$ where the exponent 
$\beta=\beta(\gamma)$ depends continuously on $\gamma$. For example, we find 
$\beta(4)=2$ and $\beta(2)=1$ \cite{SM}.


Similarly, by varying $f_{0}^{\max}$ in Fig. \ref{figABCD} one can induce 
a transition from phase $A$ to phase $C$ along the vertical line at $F=0$. 
Here we provide a concrete example of this transition by considering the attractive logarithmic
potential, $f(x)= f_0(x)= -  \frac{\alpha x}{x^2 + a^2}$, 
on the interval $[-L/2,+L/2]$, of period $L \gg a$
\cite{footnote3}. Note that in this
case the global bias in Eq. (\ref{def_bias_RTP}) vanishes, $G_L = W(-L/2) - W(L/2) = 0$, due to $f(x)$ being an odd function. To proceed, we look for 
the possible real roots of $f(x) = \pm v_0$. It is easy to verify that the four roots are given by $a\left(\mp r \pm \sqrt{r^2-1} \right)$  
where $r = \alpha/(2v_0 a)$. Clearly, if $r<1$, there is no real root -- this corresponds to phase $A$. In contrast when $r>1$ there 
are four real roots -- this corresponds to phase $C$. Thus, by tuning $r$ across the critical value $r=1$, the system can go
from phase $A$ to $C$.  

$\bullet$ For $r>1$, in phase $C$, following our general discussion before (see also Fig. \ref{fig1} bottom panel), 
there is only one region of space $[x_s, y_s]$ with $x_s = a(-r + \sqrt{r^2-1})$ and $y_s = a(r-\sqrt{r^2-1})$, where the particle gets
trapped in the stationary state, irrespective of the initial condition (since $L \gg 2 y_s$). Thus in this phase, both $V_L$
and $D_L$ vanish and the particle position is always localised (bound) at long times.

\begin{figure}[t]
\centering
\includegraphics[width=0.45\textwidth]{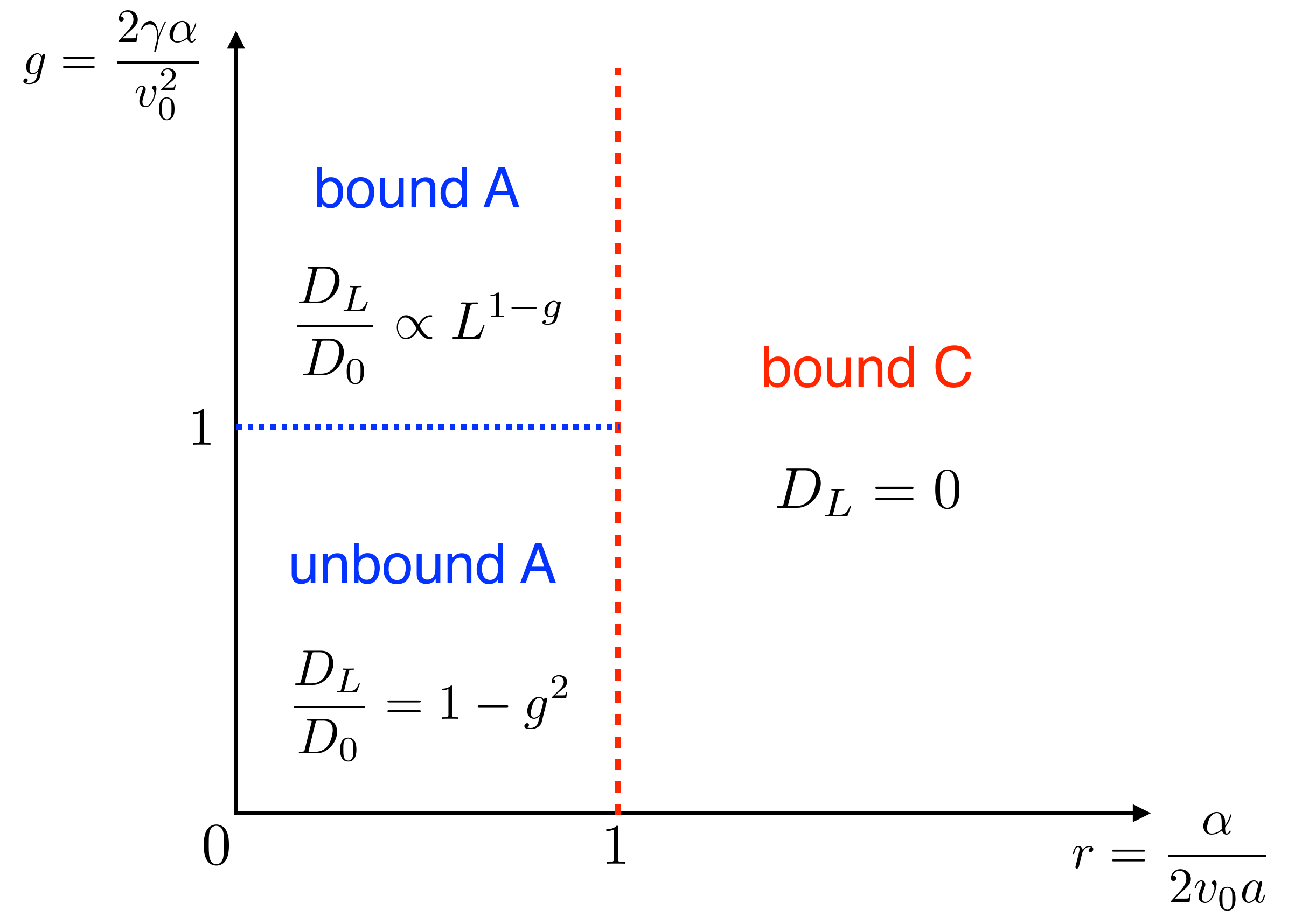}
\caption{Dynamical phase diagram of the attractive logarithmic potential model,
in terms of $r$ (activity parameter) and $g$ (potential strength),
exhibiting two binding transitions of different nature. In the
$A$ bound phase $g>1$, $r<1$, the stationary measure
$\tilde P(x)$ becomes bimodal for $g< 8 r^2$.}
\label{fig4}
\end{figure}

$\bullet$ In contrast, for $r<1$, i.e. in phase $A$, the particle position at long times may or may not be localised in the limit $L \to \infty$. This can be 
clearly seen by examining the zero-bias ($G_L=0$) stationary distribution $\tilde P(x) = \tilde A \, \Phi_+(0,x)$ where $\Phi_+$ is given in \eqref{Phipm}. It is easy to check that, for large $1 \ll |x|<L/2$, $\tilde P(x) \propto |x|^{-g}$ with the exponent $g=2 \gamma \alpha/v_0^2$. If $g>1$, the stationary distribution $\tilde P(x)$ becomes independent of $L$ in the large $L$ limit, since it can be normalised on the interval $(-\infty, + \infty)$. Thus the particle position is bound in the large $L$ limit. This can also be seen from the asymptotic behaviour of $D_L$ for large $L$, where Eq. (\ref{DL0}) predicts $D_L \sim d_r \, L^{1-g}$, with $d_r$ some $r$-dependent constant. Thus for $g>1$, the diffusion constant vanishes asymptotically for large $L$, confirming the bound state. On the other hand, if $g<1$, there is no stationary distribution in the large $L$ limit and the diffusion constant, for large $L$, approaches a constant $D_L = D_0(1-g^2)$.
This leads to the phase diagram in the $(r,g)$ plane as shown in Fig. \ref{fig4}.
For $g>1$, as $r \to 1^-$ from below (from phase $A$), the $D_L$ has an essential singularity \cite{SM}, i.e. $D_L \sim \exp(- \pi g /\sqrt{8 (1-r)})$ as $r \to 1^-$. It would be interesting to investigate the behaviour of the diffusion constant $D_L$ around the multicritical point in Fig. \ref{fig4}.

{\it Mean first passage time}. One can calculate the mean first passage time at a fixed level $X$, 
$T_\pm(x)$, for a RTP starting from $x$ in the state $\pm$. In phase $A$, for an infinite line (not assuming periodicity) assuming $W(-\infty)=+\infty$, it reads
\be
T_-(x)= \int_x^X \frac{2 \gamma dy}{v_0-f(y)} \, \int_{-\infty}^y \frac{ e^{W(y)-W(z)} \, dz}{v_0+f(z)} 
+ T_-(X)
\label{Tplus_final}
\ee
Here
$T_-(X) = \frac{1}{\gamma} + 2 
\int_{-\infty}^X \frac{e^{W(X)-W(y)} dy}{v_0 + f(y)}$
is the mean first return time to level $X$, which, for an RTP started
in the $-$ state
is non-zero, while $T_+(X)=0$. In fact the difference 
is given for general $x$ by 
\be \label{deltaTpm}
T_{-}(x) - T_{+}(x) = 2  \int_{-\infty}^x \frac{dy \, e^{W(x)-W(z)}}{v_0+f(z)} + \frac{1}{\gamma}
\ee
Note that in the diffusive limit $T_-(X) \to 0$, $T_+(x)-T_-(x) \to 0$ and one recovers the formula
given in \cite{GB1998}. These are thus purely active quantities. We have checked~\cite{SM} that the velocity
in \eqref{VelInfty} can also be obtained from the limit $\lim_{X \to +\infty} T_{\pm}(0)/X=1/V$.

All our results extend to inhomogeneous transition rates $\gamma \to \gamma(x)$ and velocity 
$v_0 \to v_0(x)$, see \cite{SM} for details. For example, in the absence of an external force $f(x)=0$,
the velocity vanishes and the diffusion constant is given by
\be \label{Dinh}
D_L = \frac{L^2 }{\left( \int_0^L dx \frac{2 \gamma(x)}{v_0(x)} \right) 
\left( \int_0^L dx \frac{1}{v_0(x)}   \right)} \;.
\ee

{\it Random landscape: velocity}.
Consider now $f(x)$ a random force where each realisation is periodic $f(x+L) = f(x)$,
but the probability distribution of $f(x)$ is independent of $x$. We restrict to the phase $A$ in the large $L$ limit. Let us define $\tilde f(x)=v_0^2 f(x)/(v_0^2- f^2(x))$, 
with $\overline{\tilde f(x)}=f_{\rm eff}$, the effective bias defined in \eqref{feff}, which we choose to be non negative
(overbars denote averages over the random force). We assume that the translational average $\langle \ldots \rangle_x$ coincides with the disorder average. This implies from \eqref{VelInfty} that the velocity
is given by 
\be
V^{-1} = \int_0^{+\infty} dz \, K(z) -  \frac{f_{\rm eff}}{v_0^2}
\ee in terms of the two point correlator 
\be \label{Kz} 
K(z)=\overline{\frac{2 \gamma v_0^2 \, e^{- (W(0)-W(z))} }{(v_0^2 - f^2(0))(v_0^2 - f^2(z))}}
\ee
There are thus two possible phases separated by a threshold force $f_c$: (i) if $\int_0^{+\infty} dz K(z) < + \infty$ there is a non zero velocity $V>0$ since the bias is positive, and (ii) $\int_0^{+\infty} dz \, K(z) = + \infty$, for which the velocity vanishes. The first case occurs for large enough $f_{\rm eff}>f_c$
since $\overline{W(0)-W(z)}=f_{\rm eff} z/D_0$. 

{\it Random landscape: anomalous diffusion}. The existence of a $V=0$ phase is a signature of anomalous
diffusion. By tuning the random force
we first consider the case $f_{\rm eff}=0$.
Consider the case where $f(x)$ is short range correlated. Then $W(x)$ performs
an unbiased random walk as a function of $x$. From~\eqref{DL0}, a good estimate,
which is also a lower bound
$\log \frac{D_L}{D_0} \geq - [\max_{x \in [0,L]} W(x) - \min_{x \in [0,L]} W(x)]  + c$,
with $c=\log \min_{x \in [0,L]}(1-\frac{f(x)}{v_0^2})$. If $W(x)$ has bounded moments,
it behaves, under rescaling, as a Brownian motion, growing typically as 
$W(x) \sim \pm \hat \sigma \sqrt{x}$, with $\hat \sigma^2 = \int_{-\infty}^{+\infty} dx \overline{\tilde f(0) \tilde f(x)}$.
This lower bound then leads to the estimate 
$\frac{1}{\sqrt L} \log D_L  \simeq - 2 \gamma \hat \sigma \omega$, where the PDF 
of $\omega>0 $ is known. The diffusion time $T_L$
on scale $L$, is thus $\log T_L = \log (L^2/D_L) \sim 2 \gamma \hat \sigma \sqrt{L}$.
This is similar to the Sinai problem \cite{Sinai}
for a passive particle, 
as noted in 
\cite{kardar}. 

Let us return to the case of non zero bias, $f_{\rm eff}>0$, not studied in \cite{kardar}.
As discussed above, $V=0$ for $f_{\rm eff}<f_c$.
The Brownian approximation for $W(x)$ allows to characterise the anomalous 
behaviour. Discarding the pre-exponential 
factors in \eqref{Kz} one obtains $V \sim f_{\rm eff}- f_c$ for $f_{\rm eff}>f_c$,
with $f_c=\frac{\gamma \sigma^2}{v_0^2}$. In the zero velocity
phase, anomalous diffusion $x \sim t^\mu$ is expected, as in the
Sinai problem \cite{Sinai}. 
Qualitatively, from Eq. \eqref{VL}, 
$1/V_L \sim \int_{x<u} e^{W(u)-W(x)} \sim e^{\Delta_m} \sim L^{\mu-1}$, 
where $\Delta_m$ is the maximum drawdown of the Brownian motion with
$\overline{\Delta_m}= \mu \log L$ where $\mu=\frac{\sigma^2}{2m}= f_{\rm eff}/f_c$
\cite{Drawdown}. 

%
%

In conclusion, we have obtained analytical expressions for the stationary measure, the velocity and the diffusion constant for a single RTP in an arbitrary 1D periodic force field with period $L$. We obtained exact results both for finite $L$ and in the large $L$ limit.  We showed that, even for a single particle, the dynamics exhibits interesting phase transitions, with power law or exponential singularities in these observables.
We also investigated \cite{SM} how the Fick's law gets modified for non-interacting RTP's subjected to a concentration gradient. It would be interesting to explore how these results are modified for interacting RTP's.

{\it Acknowledgments:} This research was partially supported by ANR grant ANR-17-CE30-0027-01 RaMaTraF.

\newpage

\begin{widetext} 

\bigskip

\bigskip

\begin{large}
\begin{center}

Supplementary Material for {\it 
Active particle in a one dimensional force field}

\end{center}
\end{large}

%
%
%
%
%
%
%
%


\bigskip

We give the principal details of the calculations described in the main text of the Letter. 

\bigskip

\medskip
\begin{center}
{\bf A. Calculation of the velocity}
\end{center}
\medskip

\noindent{\bf Stationary distribution}. As in the text we consider a RTP moving on the infinite line according to Eq. \eqref{model}, and submitted to a periodic force $f(x)=f(x+L)$. One defines $P_\pm(x,t)$ as the probability densities of the RTP to be in position $x$ at time
$t$ and in the state $\sigma(t)=\pm1$, which obey the equations \eqref{P-evol.2}, \eqref{P-evol.1}. 
One defines the periodised distributions, $\tilde P_{\pm} (x,t) = \sum_{n} P_{\pm}(x+ n L,t)$,
and the total probability $\tilde P(x,t)= \tilde P_+(x,t)+ \tilde P_-(x,t)$, as well
as the difference $\tilde Q(x,t)=\tilde P_+(x,t)-\tilde P_-(x,t)$.

Let us first obtain the stationary distributions in the phase $A$, i.e. $|f(x)|<v_0$.
Inserting the stationarity condition $\partial_t \tilde P(x)=\partial_t \tilde Q(x)=0$ in 
\eqref{tildeP} we recall that the first one is solved as $f(x) \tilde P(x) + v_0 \tilde Q(x) = J$,
where $J$ is a constant, equal to the total current. Solving for $\tilde Q(x)$ and inserting in \eqref{tildeQ} one obtains that $\tilde P(x)$ must satisfy
\be \label{eq3} 
\frac{d}{dx} [(v_0^2 - f^2(x)) \tilde P(x) + J f(x)] 
+ 2 \gamma  J - 2 \gamma f(x) \tilde P(x)  = 0 
\ee 

We then obtain the stationary distribution $\tilde P(x)$ as
\bea \label{st1} 
\tilde P(x) =  \frac{J}{v_0^2 - f^2(x)} 
\left( K e^{2 \gamma \int_0^x dy \frac{f(y)}{v_0^2 - f^2(y)}} - \int_0^x du (2 \gamma + f'(u)) e^{2 \gamma \int_u^x dy \frac{f(y)}{v_0^2 - f^2(y)}} 
 \right)
\eea 
Imposing $\tilde P(0)=\tilde P(L)$ one obtains
\bea \label{st2} 
K =  \frac{ \int_0^L du (2 \gamma + f'(u)) e^{- 2 \gamma \int_0^u dy \frac{f(y)}{v_0^2 - f^2(y)}}  }{1-
e^{- 2 \gamma \int_0^L dy \frac{f(y)}{v_0^2 - f^2(y)}} }
\eea
One can recapitulate the formula in the form
 \bea \label{st11} 
\tilde P(x) =  \frac{J}{v_0^2 - f^2(x)} \int_0^L du  (2 \gamma + f'(u))  e^{2 \gamma \int_u^x dy \frac{f(y)}{v_0^2 - f^2(y)}} 
\left( \frac{1}{1-
e^{- 2 \gamma \int_0^L dy \frac{f(y)}{v_0^2 - f^2(y)}} } - \theta(x-u)) 
 \right)
\eea 
\\

There is a useful alternative formula for $\tilde P(x)$ which does not contain derivatives of
the force. Integrating by part the above formula one obtains
\be \label{PP} 
\tilde P(x) =  J \, p(x) ~ \; {\rm with} \;\; ~ p(x) = \frac{2\gamma}{v_0^2 - f^2(x)} 
\left( \frac{\int_0^L du  \frac{v_0^2}{v_0^2-f(u)^2}  e^{2 \gamma \int_u^x dy \frac{f(y)}{v_0^2 - f^2(y)}}  }{1-
e^{- 2 \gamma \int_0^L dy \frac{f(y)}{v_0^2 - f^2(y)}} }
- \int_0^x du  \frac{v_0^2}{v_0^2-f(u)^2} e^{2 \gamma \int_u^x dy \frac{f(y)}{v_0^2 - f^2(y)}} - \frac{f(x)}{2\gamma} \right)
\ee 
which is the formula given in the text in \eqref{P0}, together with the definitions
\eqref{Phipm} for $\Phi_-(x,u)$, \eqref{AL} for $A_L$ and the definition \eqref{Wdef} of $W(x)$.
Note that although we have used here the interval $[0,L]$ as the elementary period,
a similar formula exists for any other choice of the elementary interval (such as $[-L/2,L/2]$ see
below). \\

{\it Diffusive limit for the stationary distribution}. In the limit $v_0, \gamma \to +\infty$ with $v_0^2/(2 \gamma)=D_0$ this formula becomes 
\bea
\tilde P(x) = \frac{J}{D_0} \left( \frac{ \int_0^L du e^{- \frac{1}{D_0} \int_x^u dy f(y)}  }{1-
e^{- \frac{1}{D_0}  \int_0^L dy f(y) } } - \int_0^x du e^{\frac{1}{D_0} \int_u^x dy f(y) }  \right)
\eea 
which is exactly the equation (7) in \cite{PLDV1995}. \\

\noindent{\bf Velocity}. We now calculate the velocity $V_L= \lim_{t \to +\infty} \frac{d}{dt} \int dx x P(x,t)$. We assume that $P_\pm(x,t)$ vanishes fast at $x \to \pm \infty$,
i.e. localised initial condition. Let us define
$P(x,t)= P_+(x,t)+ P_-(x,t)$, as well
as the difference $Q(x,t)=P_+(x,t)-P_-(x,t)$.
The mean instantaneous velocity of the particle (irrespective of its internal state)
is calculated as follows
\bea
&& \frac{d}{dt} \overline{x(t)} = \frac{d}{dt} \int dx x P(x,t) = - 
\int dx x \frac{d}{dx} [ f(x) P(x,t) + v_0 Q(x,t)] \\
&&= \int dx [ f(x) P(x,t) + v_0 Q(x,t)]
= \int_0^L dx [ f(x) \tilde P(x,t) + v_0 \tilde Q(x,t)] \to_{t \to +\infty} 
\int_0^L dx [ f(x) \tilde P(x) + v_0 \tilde Q(x)]  = J L \;, \nn
\eea 
where in the last line we have considered the large time limit and used
that $f(x) \tilde P(x) + v_0 \tilde Q(x) = J$ in the stationary state, where the current
$J$ is a constant. This shows Eq. \eqref{label}, i.e. that the velocity is $V_L = J L$.
Since the current $J$ is obtained by imposing the normalisation condition $\int_0^L dx \tilde P(x)=1$, we obtain $V_L$ by integrating the formula \eqref{PP} 
as $\frac{1}{V_L} = \frac{1}{L} \int_0^L dx p(x)$. This
leads to \eqref{VL} in the text.\\

We now show that, as announced in the text the sign of $V_L$ is the same as the
sign of $W(0)-W(L)$. Assume first that $\int_0^L dy \frac{f(y)}{v_0^2 - f^2(y)} > 0$. 
Since we consider phase $A$) one has $f(x)<v_0$ for all $x$. The following inequalities hold
\be
f(0) < v_0 = v_0 \frac{ 2 \gamma  \int_0^L du  \frac{f(u)}{v_0^2-f(u)^2}  
e^{- 2 \gamma \int_0^u dy \frac{f(y)}{v_0^2 - f^2(y)}}  }{1-
e^{- 2 \gamma \int_0^L dy \frac{f(y)}{v_0^2 - f^2(y)}} }
<  \frac{ 2 \gamma  \int_0^L du  \frac{v_0^2}{v_0^2-f(u)^2}  e^{2 \gamma \int_u^x dy \frac{f(y)}{v_0^2 - f^2(y)}}  }{1-
e^{- 2 \gamma \int_0^L dy \frac{f(y)}{v_0^2 - f^2(y)}} }
\ee 
where the equality in the middle is obtained by integration. This implies that
$p(0)$ given by \eqref{PP} is strictly positive. Since we know that as a probability
density $\tilde P(x) \leq 0$, it implies $J>0$, hence $V_L>0$. Similarly
if $\int_0^L dy \frac{f(y)}{v_0^2 - f^2(y)} < 0$ one has the inequalities
\be
f(0) > - v_0 = - v_0 \frac{ 2 \gamma  \int_0^L du  \frac{f(u)}{v_0^2-f(u)^2}  
e^{- 2 \gamma \int_0^u dy \frac{f(y)}{v_0^2 - f^2(y)}}  }{1-
e^{- 2 \gamma \int_0^L dy \frac{f(y)}{v_0^2 - f^2(y)}} }
>  \frac{ 2 \gamma  \int_0^L du  \frac{v_0^2}{v_0^2-f(u)^2}  e^{2 \gamma \int_u^x dy \frac{f(y)}{v_0^2 - f^2(y)}}  }{1-
e^{- 2 \gamma \int_0^L dy \frac{f(y)}{v_0^2 - f^2(y)}} }
\ee 
which implies $p(0)<0$ and hence $V_L<0$. Note that since the point $x=0$ has nothing special
one can similarly show that $\tilde P(x) >0$ for all $x$ in phase $A$.\\

\noindent{\bf Limit $\gamma \to 0$}. In the limit $\gamma \to 0$ the particle changes state only rarely.
If the particle is frozen in state $\sigma=\pm 1$, i.e. never changes state, it is easy to see that 
its velocity is
\be \label{Vsigma} 
V_\sigma = \frac{L}{\int_0^L \frac{dx}{f(x) + v_0 \sigma} }
\ee
Since the particle spends on average the same time in each state $\sigma=\pm 1$, in the limit $\gamma \to 0$ its total velocity is
\be \label{Vgamma0} 
\lim_{\gamma \to 0} V_L = \frac{1}{2} (V_+ + V_-) = \frac{L}{2 \int_0^L \frac{dx}{v_0 + f(x)} }
- \frac{L}{2 \int_0^L \frac{dx}{v_0 - f(x)} }
\ee
a formula which is valid in all phases, provided one interprets it by setting $V_\sigma=0$ if there is at least one root to the equation $f(x)=- v_0 \sigma$ with $\sigma = \pm 1$.\\

\noindent
{\bf Stationary current for a collection of independent RTP's}. Let us study a collection of independent RTP's between two reservoirs
in a segment $[0,L]$ (with no assumption here of periodicity) 
and determine the stationary current $J$ when the concentrations denoted 
$P(0)$ and $P(L)$ are fixed. We can use the formula as \eqref{st1}. The constant $K$ is then determined
by another condition. We have
\bea
&& P(0)= K \frac{J}{v_0^2 - f(0)^2} \\
&& P(L) = - \frac{J}{v_0^2 - f(L)^2} 
\left( \int_0^L du (2 \gamma + f'(u)) e^{2 \gamma \int_u^L dy \frac{f(y)}{v_0^2 - f^2(y)}} -
K e^{2 \gamma \int_0^L dy \frac{f(y)}{v_0^2 - f^2(y)}} \right) \\
&& = - \frac{J}{v_0^2 - f(L)^2} 
\left( \int_0^L du (2 \gamma + f'(u)) e^{2 \gamma \int_u^L dy \frac{f(y)}{v_0^2 - f^2(y)}} \right) +
\frac{v_0^2 - f(0)^2}{v_0^2 - f(L)^2} 
P(0) e^{2 \gamma \int_0^L dy \frac{f(y)}{v_0^2 - f^2(y)}} 
\eea
The current is then
\bea
J = -  
\frac{ (v_0^2 - f(L)^2) P(L) }{\int_0^L du (2 \gamma + f'(u)) e^{2 \gamma \int_u^L dy \frac{f(y)}{v_0^2 - f^2(y)}} } 
+ 
\frac{ (v_0^2 - f(0)^2) P(0) }{\int_0^L du (2 \gamma + f'(u)) e^{2 \gamma \int_u^0 dy \frac{f(y)}{v_0^2 - f^2(y)}} } \;.
\eea
This result, which gives back the standard formula in the diffusive limit \cite{Oshanin}, allows to study the modification of Fick's law $J \sim 1/L$, induced by an arbitrary force landscape for a RTP.

\medskip
\begin{center}
{\bf B. Calculation of the diffusion constant}
\end{center}
\medskip

\subsection{General formula} 

In addition to the functions defined in the text, to calculate the diffusion constant for an infinite
periodic medium,
we need the additional periodic functions defined as
\bea
&& \tilde S_{\pm}(x,t) = \sum_{n} (x + n L) P_{\pm}(x+ n L,t)
\eea
where $P_{\pm}(x+ n L,t)$ satisfy the evolution equations \eqref{P-evol.1}.
One easily show that the functions $S_{\pm}$ satisfy the following equations
\bea
&& \partial_t \tilde S_+ = - \partial_x [ (f(x) + v_0)  \tilde S_+]  -  \gamma \tilde S_+ + \gamma \tilde S_-  
+ (f(x) + v_0) \tilde P_+
\label{P+evol.3}\\
&& \partial_t \tilde S_- = - \partial_x [ (f(x) - v_0)  \tilde S_- ] + \gamma \tilde S_+ -  \gamma \tilde S_-  
+ (f(x) - v_0) \tilde P_- \, . 
\label{P-evol.2}
\eea
It is also convenient to define
\bea
&& \tilde S(x,t)= \tilde S_+(x,t)+ \tilde S_-(x,t) \quad , \quad \tilde R(x,t)=\tilde S_+(x,t)-\tilde S_-(x,t) 
\eea
which satisfy
\bea \label{RS}
&& \partial_t \tilde S = - \frac{d}{dx} [ f(x) \tilde S + v_0 \tilde R]  + f(x) \tilde P + v_0 \tilde Q \\
&& \partial_t \tilde R = - \frac{d}{dx} [ f(x) \tilde R + v_0 \tilde S] - 2 \gamma \tilde R 
+ f(x) \tilde Q + v_0 \tilde P \nonumber
\eea
where $\tilde P$ and $\tilde Q$ are defined in the text. One expects that in the large time limit,
as is the case in the diffusive problem~\cite{DerridaLong}
\bea
&& \tilde P(x,t) \to \tilde P(x) \quad , \quad \tilde Q(x,t) \to \tilde Q(x)  \\
&& \tilde S(x,t) \to  a(x) t + s(x) \quad , \quad \tilde R(x,t) \to b(x) t + r(x) 
\eea 
Note that all these functions are periodic in $x$ of period $L$.

Injecting this form into \eqref{RS} and collecting the terms proportional to 
$t$, we see that $a(x)$ and $b(x)$ satisfy the same equation 
as the stationary solutions $\tilde P(x)$ and $\tilde Q(x)$ 
given in the text in \eqref{tildeP}. Hence we set
\be
a(x) = A \, \tilde P(x) \quad , \quad b(x) = A \, \tilde Q(x) 
\ee
where $A$ is for now un unknown constant.

The equations which determine $s(x)$ and $r(x)$ are then
\bea \label{rs0} 
&& A \tilde P = - \frac{d}{dx} [ f(x) s + v_0 r]  + f(x) \tilde P + v_0 \tilde Q \\
&& A \tilde Q = - \frac{d}{dx} [ f(x) r + v_0 s] - 2 \gamma r 
+ f(x) \tilde Q + v_0 \tilde P  \;. \label{rs00} 
\eea
We will study these equations below. If one knows the solution, one can
obtain the diffusion constant as follows. One can write the instantaneous velocity,
and its large time limit as
\bea
\overline{x(t)}  = \int_{-\infty}^{+\infty} dx \, x \, P(x,t) = 
\int_0^L dx \tilde S(x,t) 
\to t \int_0^L dx \, a(x) + \int_0^L dx \, s(x) 
\eea 
where we have used that $\tilde S(x,t) = \sum_{n} (x + n L) P(x+ n L,t)$.
The above equations allows to identify both $A$ and $J L$ with the velocity $V$, 
\bea
J L = V = \int_0^L dx a(x) = A \int_0^L dx \, \tilde P(x) = A
\eea 
Note that $\int_0^L dx s(x)$ is a shift at large time. Next, we have
\bea
&& \frac{d}{dt} \overline{x(t)^2} = \int dx x^2 \partial_t P(x,t) =
- \int dx x^2 \partial_x ( f(x) P + v_0 Q ) 
\\
&& = 2 \int dx x [ f(x) P + v_0 Q] =
2 \int_0^L dx [ f(x) \tilde S(x,t) + v_0 \tilde R(x,t) ]
\\
&& \to 2 \int_0^L dx [ f(x) a(x) + v_0 b(x) ] t + 2 \int dx [ f(x) s(x) + v_0 r(x) ] \\
&& = 2 A \int_0^L dx [ f(x) \tilde P(x) + v_0 \tilde Q(x) ] t + 2 \int dx [ f(x) s(x) + v_0 r(x) ]
= 2 A^2 t + 2 \int dx [ f(x) s(x) + v_0 r(x) ]
\eea 
The diffusion constant $D_L$ is now given by
\bea \label{D1} 
D_L = \frac{1}{2} \lim_{t \to +\infty} \left(\frac{d}{dt} \overline{x(t)^2}  - \frac{d}{dt} \overline{x(t)}^2 \right)
= \int dx [ f(x) s(x) + v_0 r(x) ] - A \int_0^L dx s(x)
\eea 
hence it requires the knowledge of $s(x)$ and $r(x)$.

\subsection{Diffusion constant in phase $A$, and in the absence of a bias}

Let us first study the phase $A$, i.e. $|f(x)|<v_0$, with zero bias, i.e. $W(L)=W(0)$ i.e. 
$\int_0^L dy \frac{f(y)}{v_0^2 - f^2(y)}  = 0$, for which $V_L = J L=A=0$. 
The stationary distribution is given by
\bea
\tilde P(x) =  \tilde A \frac{v_0^2}{v_0^2 - f^2(x)} 
e^{2 \gamma \int_0^x dy \frac{f(y)}{v_0^2 - f^2(y)}} \quad , \quad 
\frac{1}{\tilde A} = \int_0^L dx \frac{v_0^2}{v_0^2 - f^2(x)} 
e^{2 \gamma \int_0^x dy \frac{f(y)}{v_0^2 - f^2(y)}}  \label{At}
\eea 
where the equation for $\tilde A$ is determined by the normalisation condition
$\int_0^L dx \tilde P(x)=1$. Using that $f(x) \tilde P(x) + v_0 \tilde Q(x)=J=0$,
the equations \eqref{rs0}, \eqref{rs00}, simplify into
\bea \label{rs2} 
&& 0 = - \frac{d}{dx} [ f(x) s + v_0 r]   \\
&& 0 = - \frac{d}{dx} [ f(x) r + v_0 s] - 2 \gamma r
+ \frac{v_0^2-f^2(x)}{v_0} \tilde P \label{rs20} 
\eea
The first equation gives
\be \label{rrr} 
f(x) s(x) + v_0 r(x) = c 
\ee 
where $c$ is a constant. From $c$ one obtains from \eqref{D1} (with $A=0$)
the diffusion constant as $D= c L$. Solving for $s(x)$ and inserting 
in \eqref{rs20} we obtain an equation for $r(x)$
\be \label{eqq} 
0 = - \frac{d}{dx} [ \frac{f^2(x) - v_0^2}{f(x)} r(x)]  + c v_0 \frac{f'(x)}{f^2(x)}  - 2 \gamma r(x)
+ \tilde A v_0 
e^{2 \gamma \int_0^x dy \frac{f(y)}{v_0^2 - f^2(y)}} 
\ee
Let us define the auxiliary function $\tilde r(x)$ as
\be \label{rtilde} 
r(x)=\tilde r(x) \frac{f(x)}{v_0^2 - f^2(x)} 
e^{2 \gamma \int_0^x dy \frac{f(y)}{v_0^2 - f^2(y)}} \;.
\ee
Note that since $r(x)$ and $f(x)$ are periodic and $W(L)=W(0)$,
the function $\tilde r(x)$ is also periodic of period $L$.
From \eqref{eqq} we see that it satisfies 
\bea
\tilde r'(x) = c v_0 \frac{d}{dx}  \left[\frac{1}{f(x)}\right] e^{- 2 \gamma \int_0^x dy \frac{f(y)}{v_0^2 - f^2(y)}}  
- \tilde A \, v_0 \;.
\eea 
We find by integrating from $0$ to $L$ and using $\tilde r(0)=\tilde r(L)$ 
\be
c = \frac{D_L}{L} = \frac{\tilde A \, L}{\int_0^L dx \frac{d}{dx}\left[\frac{1}{f(x)}\right] e^{- 2 \gamma \int_0^x dy \frac{f(y)}{v_0^2 - f^2(y)}} }
\ee 
which can be integrated by parts (with a vanishing boundary term since the bias is zero). Using the formula \eqref{At} for $\tilde A$ we finally obtain
\bea \label{DLSM} 
D_L = \frac{L^2}{2 \gamma \left( \int_0^L dx \frac{1}{v_0^2- f^2(x)}  
e^{- 2 \gamma \int_0^x dy \frac{f(y)}{v_0^2 - f^2(y)}} \right) 
\left( \int_0^L dx \frac{v_0^2}{v_0^2- f^2(x)}  
e^{2 \gamma \int_0^x dy \frac{f(y)}{v_0^2 - f^2(y)}} \right)}
\eea 
This gives the formula \eqref{DL0} of the text, where we used $D_0= \frac{v_0^2}{2 \gamma}$.
For $f(x)=0$ one recovers $D_L= D_0$. In the diffusive limit, $v_0, \gamma \to \infty$ with 
$D_0$ fixed, this formula becomes
\bea
D_L = D_0 \frac{L^2}{\left( \int_0^L dx 
e^{- \frac{1}{D_0}  \int_0^x dy f(y) } \right)
\left( \int_0^L dx 
e^{\frac{1}{D_0}  \int_0^x dy f(y) }\right)}
\eea 
which is the standard formula in the diffusive case.\\

\subsection{Diffusion constant in phase $A$, and in the presence of a bias}

Let us consider now the case where the bias is non zero, i.e. $V_L \neq 0$. Let us recall
the expression \eqref{D1} for the diffusion constant where we use that $A=V_L$, namely
\bea \label{DDD} 
D_L =  \int_0^L dx [ f(x) s(x) + v_0 r(x) ] - V_L \int_0^L dx s(x)
\eea 
Let us also recall that the equations which determine $s(x)$ and $r(x)$ are
\bea \label{vv}
&& V_L \tilde P = - \frac{d}{dx} [ f(x) s + v_0 r]  + f(x) \tilde P + v_0 \tilde Q \\
&& V_L \tilde Q = - \frac{d}{dx} [ f(x) r + v_0 s] - 2 \gamma r 
+ f(x) \tilde Q + v_0 \tilde P \nonumber 
\eea
and that we can use that $f(x) \tilde P + v_0 \tilde Q = J$. We should also remember that
$r(x)$ and $s(x)$ are periodic of period $L$. The first equation gives
\be \label{cons} 
f(x) s(x) + v_0 r(x) = J \int_0^x dy (1 - L \tilde P(y)) + c
\ee 
where $c$ is an integration constant. We will see below that its value is
immaterial for calculating $D_L$, but for now we keep it. We can thus
express $r(x)$ as a function of $s(x)$ and insert its expression in the second equation in 
\eqref{vv}. Replacing $V_L = J L$, we obtain the following equation for $s(x)$
%
%
\bea 
&& 0 = - \frac{d}{dx} [(v_0^2 - f^2(x))  s(x)  ] + 2 \gamma f(x) s(x) - c (f'(x) + 2 \gamma) + B(x) \\
&& B(x) = 
- \frac{d}{dx} [ J f(x)  \int_0^x dy (1 - L \tilde P(y))) ) ]  \label{By} \\
&& - 2 \gamma J \int_0^x dy (1 - L \tilde P(y))
+ (v_0^2 - f^2(x) + J L f(x)) \tilde P(x) + f(x) J - L J^2 
\eea 
To solve this equation one defines the auxiliary function $\tilde s(x)$ via 
\be
s(x)=\tilde s(x) \frac{1}{v_0^2 - f^2(x)} 
e^{2 \gamma \int_0^x dy \frac{f(y)}{v_0^2 - f^2(y)}}
\ee
One finds that it satisfies the equation 
\be
 \tilde s'(x) = (B(x) - c (f'(x) + 2 \gamma) ) e^{- 2 \gamma \int_0^x dy \frac{f(y)}{v_0^2 - f^2(y)}} 
\ee
which we integrate as follows
\be
\tilde s(x) = \tilde s(0) + \int_0^x dy (B(y) - c (f'(y) + 2 \gamma) ) e^{- 2 \gamma \int_0^y du \frac{f(u)}{v_0^2 - f(u)^2}}
\ee 
Hence we now have a second unknown integration constant, $\tilde s(0)$. 
It can be fixed however from the periodicity of $s(x)$. Writing
$s(0)=s(L)$ we obtain the condition
\bea
\tilde s(0) (1 - e^{2 \gamma \int_0^L dy \frac{f(y)}{v_0^2 - f^2(y)}}) =
 \int_0^L dy (B(y)  -  c (f'(y) + 2 \gamma) ) e^{- 2 \gamma \int_L^y du \frac{f(u)}{v_0^2 - f(u)^2}} \;.
\eea
Substituting we write the final result for $s(x)$
\bea
&& s(x) = \frac{1}{v_0^2 - f^2(x)} 
e^{2 \gamma \int_0^x dy \frac{f(y)}{v_0^2 - f^2(y)}}
\bigg( \frac{1}{1 - e^{2 \gamma \int_0^L dy \frac{f(y)}{v_0^2 - f^2(y)}}} \int_0^L dy (B(y)  -  c (f'(y) + 2 \gamma) ) e^{- 2 \gamma \int_L^y du \frac{f(u)}{v_0^2 - f(u)^2}} 
 \\
 && ~~~~~~~~~~~~~~~~~~~~~~~~~~
 + \int_0^x dy (B(y) - c (f'(y) + 2 \gamma) ) e^{- 2 \gamma \int_0^y du \frac{f(u)}{v_0^2 - f(u)^2}} \bigg) \nn
\eea 
From \eqref{DDD} and \eqref{cons} the diffusion constant is given by (upon insertion of
the result for $s(x)$ and various manipulations) 
\bea
&& D_L = J \int_0^L dx \int_0^x dy (1 - L \tilde P(y)) + c L - J L  \int_0^L dx s(x) \\
&& = J \int_0^L dx \int_0^x dy (1 - L \tilde P(y)) + c L \\
&& - J L  
\int_0^L dx \int_0^L dy \frac{e^{ 2 \gamma \int_y^x du \frac{f(u)}{v_0^2 - f(u)^2} }}{v_0^2 - f^2(x)} (B(y) - c (f'(y) + 2 \gamma) ) 
[\theta(x-y) - \frac{1}{1 - e^{-2 \gamma \int_0^L du \frac{f(u)}{v_0^2 - f(u)^2}}}]
\eea
Now we note, remarkably, that the terms proportional to $c$ cancel because of the normalisation
condition $\int_0^L dx \tilde P(x)=1$ applied on Eq. \eqref{st11}. Hence we
simply obtain 
\bea \label{DL1} 
D_L = J \int_0^L dx \int_0^x dy (1 - L \tilde P(y)) - J L  
\int_0^L dx \int_0^L dy \frac{e^{ 2 \gamma \int_y^x du \frac{f(u)}{v_0^2 - f(u)^2} }}{v_0^2 - f^2(x)} B(y)  
[\theta(x-y) - \frac{1}{1 - e^{-2 \gamma \int_0^L du \frac{f(u)}{v_0^2 - f(u)^2}}}]
\eea 
where $B(y)$ is given in \eqref{By} which we rewrite as
\be \label{DL2} 
B(x) = - J (2 \gamma + f'(x))  \int_0^x dy (1 - L \tilde P(y)) 
+ (v_0^2 - f^2(x) + 2 J L f(x)) \tilde P(x) - L \, J^2 
\ee
The final result for the diffusion constant in the presence of a drift is given 
by substituting the expression \eqref{PP} (equivalently \eqref{st11}) for $\tilde P(x)$ in Eqs. \eqref{DL1}, \eqref{DL2}. Although it is a complicated formula, it is valid for any force landscape such that $|f(x)|<v_0$ (phase $A$). 

We can now check that in the limit of zero bias this formula crosses over smoothly to 
formula \eqref{DLSM} (i.e. \eqref{DL0} in the text) for the zero bias diffusion constant.
First it is easy to see from Eq. \eqref{PP} that the stationary distribution $\tilde P(x)$ 
converges to the one given by formula \eqref{At} in the zero bias limit where
both $J$ and $\int_0^L du \frac{f(u)}{v_0^2 - f(u)^2}$ tend to zero. This can 
be seen since in that limit the r.h.s. in \eqref{DLSM} is dominated by the
first term. The normalisation condition $\int_0^L dx \tilde P(x)=1$
in Eq. \eqref{DLSM} further implies that, in the 
limit $J \to 0$ the following ratio goes to a constant
\be
\frac{J}{ 1- e^{-2 \gamma \int_0^L du \frac{f(u)}{v_0^2 - f(u)^2}}} \to \frac{D^{\rm zb}_L}{L^2}
\ee 
where $D^{\rm zb}_L$ is the zero bias diffusion constant given by Eq. \eqref{DLSM}.
This is nothing but the Einstein relation [see Eq. (\ref{E}) in the main text], i.e., $V_L \simeq D^{\rm zb}_L \frac{G_L}{L} =  D^{\rm zb}_L \frac{1}{L}(W(0)-W(L))$ in the small bias limit, where we have used $G_L = W(0) - W(L)$. We note that when $J \to 0$ one has
\be
B(y) \simeq (v_0^2 - f^2(y)) \tilde P(y) \simeq \tilde A \, v_0^2 e^{ 2 \gamma 
\int_0^y du \frac{f(u)}{v_0^2 - f(u)^2}} \;,
\ee
where $\tilde A$ is given in \eqref{At}. Substituting in \eqref{DL1} one sees
that the integrand in the last term (which is the only remaining term
in the limit $J \to 0$) does not depend on $y$. This produces a factor
$L^2$ and one can check that the remaining integral over $x$ cancels
exactly the factor $\tilde A$, with the result $D_L \to D^{\rm zb}_L$
when $J \to 0$. 
\medskip
\begin{center}
{\bf C. Piecewise linear force model}
\end{center}
\medskip

We provide additional information on the study of the model $v_0=1$ and 
$f(x)=4 |x-\frac{1}{2}| - 1 + F$ in the interval $x \in [0,1]$, i.e. we set $L=1$ here. For $\gamma=4$ and
$F=1$ this model was studied in the text. The stationary measure is 
given in \eqref{plusieurs}. Here we study other values of $\gamma$ and $F$.\\

\noindent{\bf The case $\gamma=2$, $F=1$}. Consider first the case $\gamma=2$ and $F=1$ where more explicit formulae can be given. We find, by solving explicitly \eqref{eq3}
in each subinterval where $f(x)$ is linear
\bea \label{plusieurs2}
&& \tilde P(x) = \frac{c_1}{\sqrt{(1-4 x)(3-4 x)} } \quad , \quad 0<x<1/4 \\
&& \tilde P(x) = \frac{c_2}{\sqrt{(4 x-1)(3-4 x)} } \quad , \quad 1/4<x<1/2 \\
&& \tilde P(x) = \frac{2 J \left(\sqrt{(3-4 x) (4 x-1)} (1-2 x)+\sin
   ^{-1}\left(\sqrt{\frac{3}{2}-2 x}\right)\right)}{((3-4 x)
   (4 x-1))^{3/2}} \quad , \quad 1/2 < x < 3/4 \\
&& \tilde P(x) = \frac{J \left(2 (2 x-1) \sqrt{16 x^2-16
   x+3}-\log \left(\sqrt{16 x^2-16 x+3}+4
   x-2\right)\right)}{((4 x-3) (4
   x-1))^{3/2}} \quad , \quad 3/4<x<1 
\eea 
where $c_1$ and $c_2$ are integration constants. The singularity at the unstable fixed point (see text)
$x=3/4$ is only apparent on these formula. Note that the general solution on the interval $x \in [1/2,1]$ involves an additional term multiplied by an integration constant. This term however has a non-integrable divergence at $x=3/4$, hence the coefficient can be set to zero.
There is no change at $x=3/4$ since we have chosen the integration constant to be zero 
leading to the third line. In fact the last two lines in \eqref{plusieurs2} are analytical continuations of each other.
This is a general feature of the problem, valid for any $\gamma$. Imposing the continuity of 
$\tilde P(x)$ at $x=1/2$, the periodicity $\tilde P(0)=\tilde P(1)$, and the normalisation
$\int_0^1 dx \, \tilde P(x)=1$ we can determine all the unknown constants $c_1$, $c_2$ and $J$ as
\bea
&& c_1=\frac{1}{3} J \left(2 \sqrt{3}-\log
   \left(2+\sqrt{3}\right)\right) \\
&&   c_2= \frac{\pi  J}{2} \\
&& 1= \frac{1}{48} J \left(3 \pi ^2+4 \left(4
   \sqrt{3}+\log
   \left(2-\sqrt{3}\right)\right) \cosh
   ^{-1}(2)\right)
\eea which, using $V_L = J L$ with $L=1$, leads to the result for the velocity
\bea
V_L = \frac{48}{3 \pi ^2+4 \left(4 \sqrt{3}+\log
   \left(2-\sqrt{3}\right)\right) \cosh^{-1}(2)} = 0.811251\ldots
\eea  
with $\cosh^{-1}(2) = \log(2 + \sqrt{3})$.\\

\noindent{\bf The case of arbitrary $\gamma$ for $F=1$.} Performing the same steps
as before we obtain the stationary measure as
\bea
&& \tilde P(x) = c_1 \left((1-4 x) (3-4 x)\right)^{\frac{\gamma  }{4}-1}
+ \frac{(\gamma -2) J \, _2F_1\left(1,2-\frac{\gamma
   }{2};2-\frac{\gamma }{4};2 x-\frac{1}{2}\right)}{\gamma -4}   \quad , \quad 0<x<1/4 \\
&& \tilde P(x) = c_2 \left((4 x-1)(3-4 x) \right)^{\frac{\gamma}{4}-1}
+ 
\frac{(\gamma -2) J \, _2F_1\left(1,2-\frac{\gamma
   }{2};2-\frac{\gamma }{4};2 x-\frac{1}{2}\right)}{\gamma -4}
    \quad , \quad 1/4<x<1/2 \\
&& \tilde P(x) = \frac{(\gamma +2) J \, _2F_1\left(1,\frac{\gamma
   +4}{2};\frac{\gamma +8}{4};\frac{3}{2}-2 x\right)}{\gamma
   +4}
 \quad , \quad 1/2<x<1\;,
\eea
where $_2F_1(a,b;c;z)$ denotes the Gauss hypergeometric function. 
\begin{figure}[t]
\includegraphics[width=0.5\linewidth]{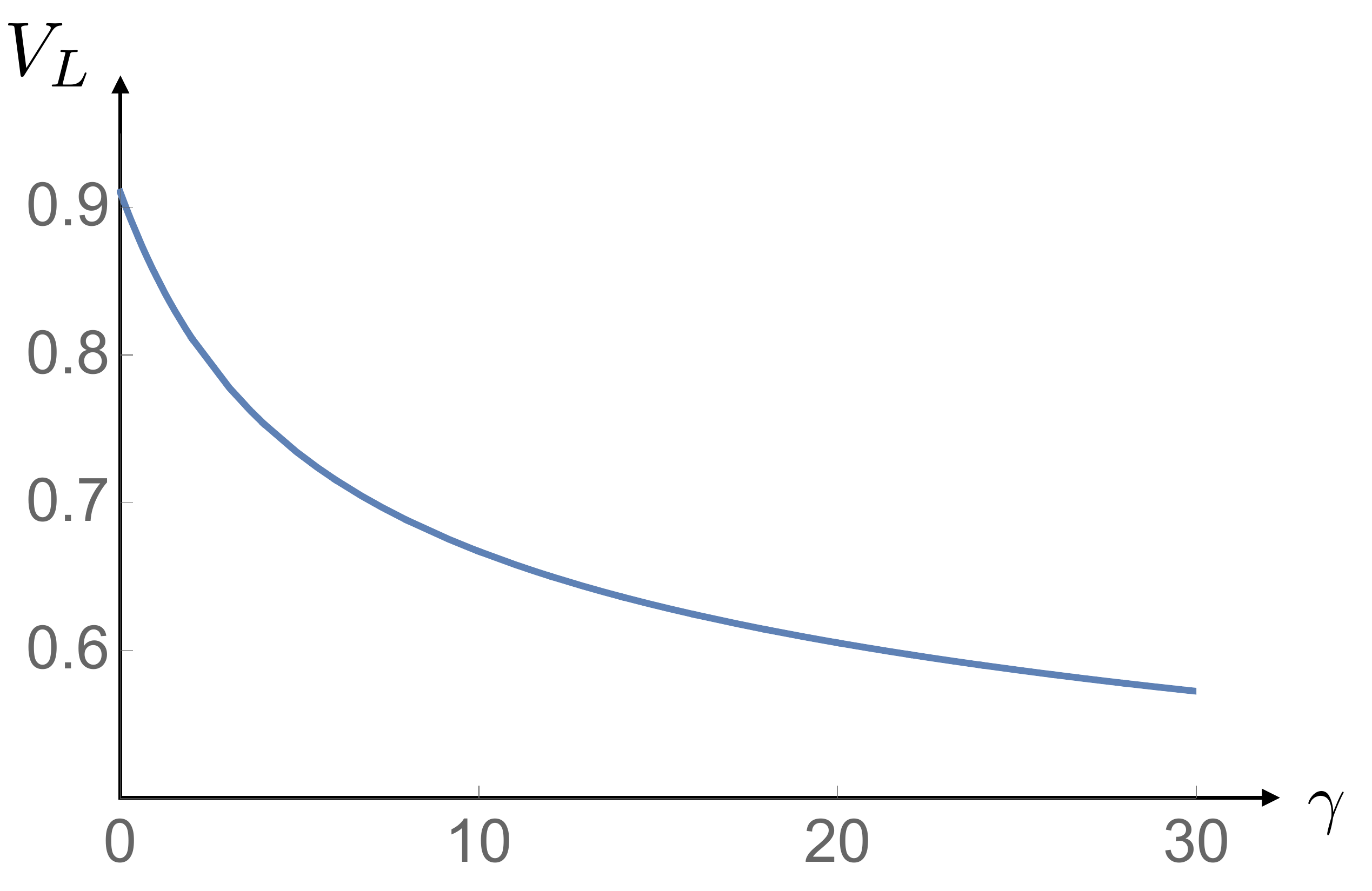}
\caption{Plot of the velocity $V_L$ for the potential $f(x) = 4|x-1/2|$ vs $\gamma$, as given by the exact formula in Eq. (\ref{VL_gamma}). In particular, in the limit $\gamma \to 0$, one has $V_L = 1/\log 3 = 0.910239 \ldots$ [see Eq. (\ref{limgamma0})].}\label{Fig_VL_gamma}
\end{figure}
Proceeding as before we determine the integration constants $c_1,c_2,J$ and finally obtain
the velocity for any $\gamma>0$ as
\bea \label{VL_gamma}
&& \frac{1}{V_L} = \frac{(\gamma -2) \left(\, _3F_2\left(1,1,2-\frac{\gamma
   }{2};2,2-\frac{\gamma }{4};-\frac{1}{2}\right)+\,
   _3F_2\left(1,1,2-\frac{\gamma }{2};2,2-\frac{\gamma
   }{4};\frac{1}{2}\right)\right)}{4 (\gamma
   -4)}
   \\
   && +\frac{(\gamma +2) \left(\, _3F_2\left(1,1,\frac{\gamma
   }{2}+2;2,\frac{\gamma }{4}+2;-\frac{1}{2}\right)+\,
   _3F_2\left(1,1,\frac{\gamma }{2}+2;2,\frac{\gamma
   }{4}+2;\frac{1}{2}\right)\right)}{4 (\gamma
   +4)} \nn \\
   && +\frac{1}{8} \left(\frac{\cos \left(\frac{\pi  \gamma
   }{4}\right) \Gamma \left(\frac{1}{2}-\frac{\gamma
   }{4}\right) \Gamma \left(\frac{\gamma }{4}+1\right) \Gamma
   \left(\frac{\gamma }{4}\right)}{\Gamma \left(\frac{\gamma
   +2}{4}\right)}-\pi  \cot \left(\frac{\pi  \gamma
   }{4}\right)\right) \nn \\
   && + \frac{1}{\gamma} \left(\frac{3}{2}\right)^{1-\frac{
   \gamma }{4}} \left(\frac{(\gamma +2) \,
   _2F_1\left(1,\frac{\gamma +4}{2};\frac{\gamma
   +8}{4};-\frac{1}{2}\right)}{\gamma +4}-\frac{(\gamma -2) \,
   _2F_1\left(1,2-\frac{\gamma }{2};2-\frac{\gamma
   }{4};-\frac{1}{2}\right)}{\gamma -4}\right) \,
   _2F_1\left(1-\frac{\gamma }{4},\frac{\gamma
   }{4};\frac{\gamma +4}{4};-\frac{1}{2}\right) \nn \;,
\eea
where $_3F_2(a_1, a_2, a_3;b_1, b_2;z)$ denotes a generalised hypergeometric function. This formula is plotted in Fig. \ref{Fig_VL_gamma}. We see that $V_L$ is a decreasing function of $\gamma$. 
In the limit $\gamma \to 0$ we can use the prediction given in  \eqref{Vsigma}, \eqref{Vgamma0},
i.e. 
\be\label{limgamma0}
\lim_{\gamma \to 0} V_L = \frac{1}{2 \int_0^L \frac{dx}{v_0 + f(x)} } =\frac{1}{2 \int_0^1 \frac{dx}{1+ 4 |x - \frac{1}{2}|} } = \frac{1}{\log 3} = 0.910239\ldots,
\ee
which is found to be in very good agreement with the numerical evaluation of $V_L$ (see also Fig. \ref{Fig_VL_gamma}) as well as with the Taylor expansion
$V_L = 0.910239\ldots - 0.0654649 \ldots \gamma + O(\gamma^2)$ performed with Mathematica.\\

\noindent{\bf The case $\gamma=4$ and $0 < F < 2$.} Consider $\gamma=4$ and vary the external force $F$ while remaining in phase $B$ (see the upper panel of Fig. \ref{fig1}), i.e. $0 < F < 2$. 
The critical points are $x_s=\frac{F}{4}$ and $x_u=\frac{4-F}{4}$. We find that the stationary measure is
given by 
\bea \label{plusieurs5}
&& \tilde P(x) = c_1 + \frac{1}{2} J (\log (F-4 x+2)-\log (F-4 x)) \quad , \quad 0<x<\frac{F}{4} \;, \\
&& \tilde P(x) = c_2 + \frac{1}{2} J (\log (F-4 x+2)-\log (4 x- F)) \quad , \quad \frac{F}{4}<x<1/2 \;, \\
&& \tilde P(x) 
= \frac{J (F+4 x-1)}{(F+4 x-2)^2}
\quad , \quad 1/2<x<1 \;.
\eea 
One finds the constants $c_1,c_2,J$ from the continuity of $\tilde P(x)$ at $x=1/2$, the periodicity $\tilde P(0)=\tilde P(1)$ and the
normalisation, leading to the following formula for the velocity
   \be \label{VL_gamma4}
V_L = \frac{2 F^2 (F+2)^2}{2 ((F-1) (F+5)+7)- F^2 (F+2)^2 \log F+ F^2 (F+2)^2 \log (F+2)} \;.
   \ee
This formula for $V_L$ vs $F$ for $\gamma=4$ is plotted in the left panel of Fig. \ref{Fig_VL_gamma2_4}.    
From this formula (\ref{VL_gamma4}), one obtains that the velocity vanishes quadratically as $F \to 0^+$ as
\be
V_L = 2 F^2-2 F^3+F^4 \left(2 \log (F)+\frac{7}{2}-\log
   (4)\right)+O\left(F^5\right) \;,
\ee 
with some subdominant logarithmic singularities.
 \\
 \begin{figure}
 \includegraphics[width = 0.8\linewidth]{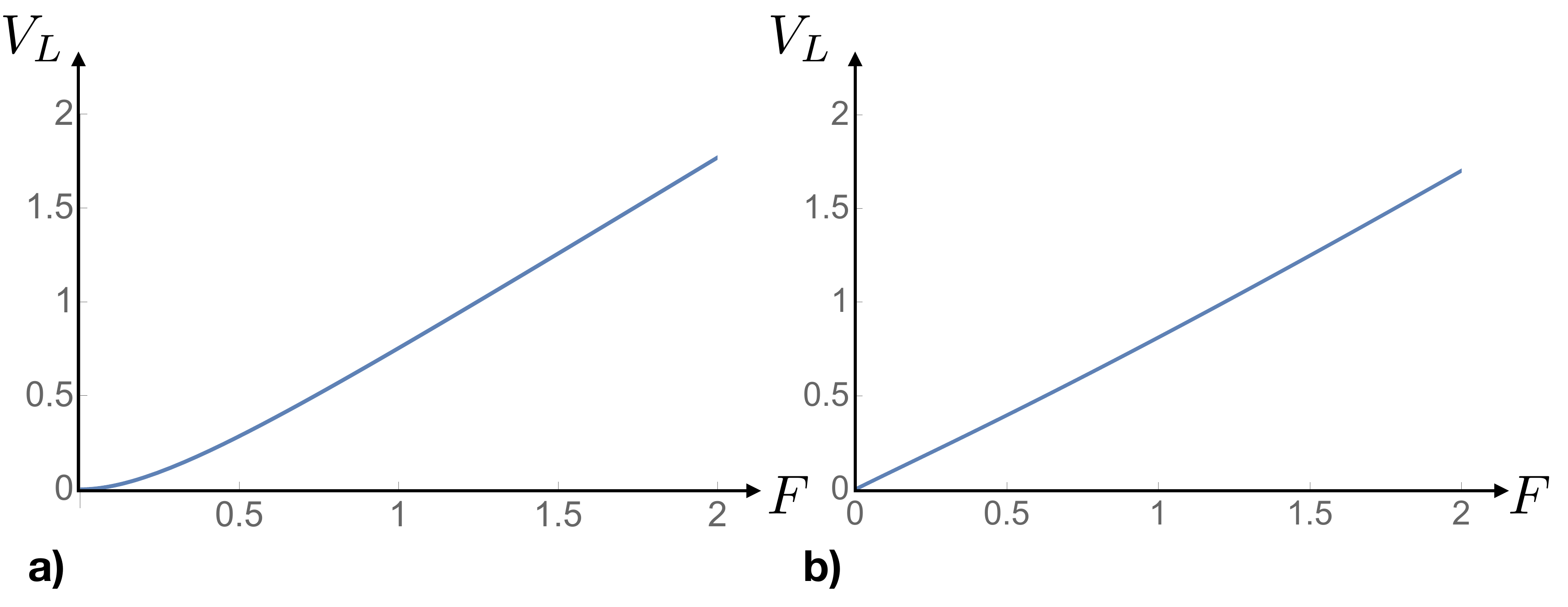}
 \caption{Plot of $V_L$ for the potential $f(x) = 4 |x-1/2| + F-1$ as a function of $F \in [0,2]$ for two different values of $\gamma$: {\bf a)} for $\gamma = 4$, as given in Eq. (\ref{VL_gamma4}) and {\bf b)} for $\gamma =2$ as given in Eq. (\ref{VL_gamma2}). While the two formulae in Eq. (\ref{VL_gamma4}) and (\ref{VL_gamma2}) are quite different, the two curves shown in panel a) and panel b) look very similar on that scale. However, they actually differ by their behaviour near $F=0$, with $V_L \propto F^2$ for $\gamma=4$ (left panel) while $V_L \propto F$ for $\gamma=2$ (right panel).}\label{Fig_VL_gamma2_4}
 \end{figure}


\noindent{\bf The case $\gamma=2$ and $0 < F < 2$.} We now find the stationary measure
\bea \label{plusieurs2}
&& \tilde P(x) = \frac{c_1}{\sqrt{(F-4 x)(2+F-4 x)} } \quad , \quad 0<x<\frac{F}{4} \;, \\
&& \tilde P(x) = \frac{c_2}{\sqrt{(4 x-F)(2+F-4 x)} } \quad , \quad \frac{F}{4}<x<1/2 \;, \\
&& \tilde P(x) =  \frac{(\gamma +2) J \, _2F_1\left(1,\frac{\gamma
   +4}{2};\frac{\gamma +8}{4}; 2 (\frac{4 - F}{4}-x) \right)}{\gamma
   +4}|_{\gamma=2} \quad , \quad 1/2<x<1 \;,
\eea 
which leads to the formula for the velocity
%
\bea \label{VL_gamma2}
&&  \frac{1}{V_L} = \frac{\sqrt{F (F+2)}}{4 F^{3/2} (F+2)^{3/2}} \bigg(2 \left(\sqrt{F} \sqrt{F+2}
   (F+1)+\log \left(F-\sqrt{F (F+2)}+1\right)\right) \sinh
   ^{-1}\left(\frac{\sqrt{F}}{\sqrt{2}}\right) \\
   && -\sqrt{F} (F+1)
   \sqrt{F+2} \log \left(F-\sqrt{F
   (F+2)}+1\right)\bigg)
   \\
   && 
   +\frac{4
   \left((F-1) \sqrt{-(F-2) F}-\sin
   ^{-1}\left(\sqrt{1-\frac{F}{2}}\right)\right) \sin
   ^{-1}\left(\sqrt{1-\frac{F}{2}}\right)}{(F-2) F} \nn \;.
\eea 
This formula for $V_L$ vs $F$ for $\gamma=2$ is plotted in the right panel of Fig. \ref{Fig_VL_gamma2_4}. 
Around $F=1$ it reproduces the result given above, i.e. one finds $V_L \simeq 0.811251\ldots+0.854383\ldots (F-1)$. Now we find that it vanishes linearly as $F \to 0^+$ 
\be
V_L = \frac{8 F}{\pi ^2}-\frac{4 F^2}{\pi ^2}+\frac{64 \sqrt{2}
   F^{5/2}}{3 \pi ^3}-\frac{64 F^3}{\pi ^4}-\frac{208 \sqrt{2}
   F^{7/2}}{15 \pi ^3}+\frac{1600 F^4}{9 \pi
   ^4}+O\left(F^{9/2}\right) \;,
\ee
with subdominant half integer powers.

%


%
%
%

\medskip
\begin{center}
{\bf D. RTP in a logarithmic potential}
\end{center}
\medskip


\begin{figure}
\includegraphics[width = 0.6\linewidth]{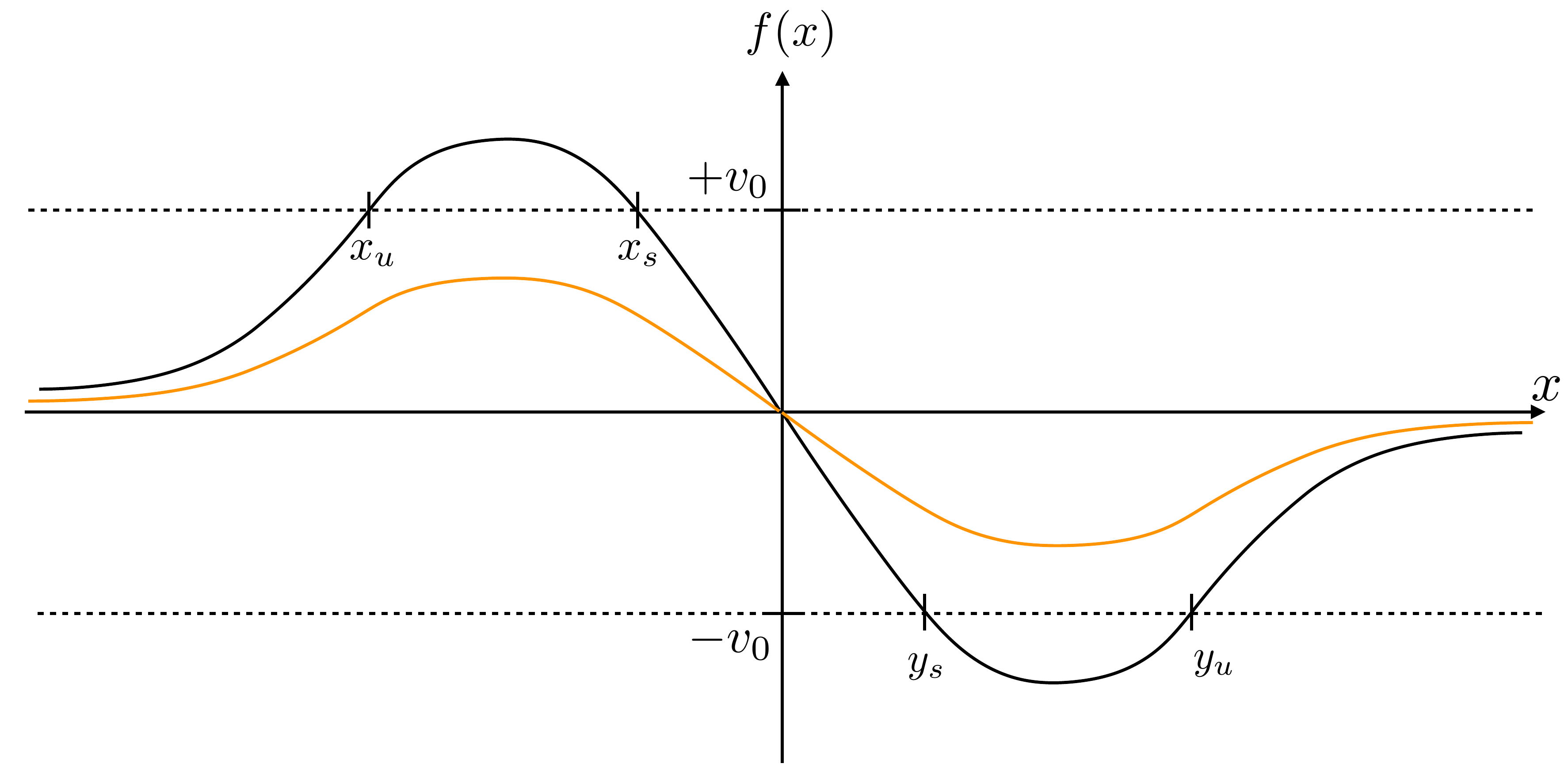}
\caption{Plot of the force $f(x) =  - \frac{\alpha x}{x^2 + a^2}$ vs $x$ with $a=1$ and for two different values of the parameter $r$: $r<1$ (orange solid line), corresponding to the phase $A$ and $r>1$ (black solid line) corresponding to the phase $C$. In the latter case, $x_s$ and $x_u$ denote respectively the stable and unstable fixed points in the state $\sigma = -1$, while $y_s$ and $y_u$ denote their counterpart in the state $\sigma = +1$ (see Fig. \ref{fig1} in the main text). In phase $C$ (black solid line), the stationary measure $\tilde P(x)$ is supported on the interval $[x_s, y_s]$.}\label{Fig_log}
\end{figure}

\noindent
{\bf Stationary distribution}. Consider, as in the text, the attractive logarithmic potential $U(x)=\frac{\alpha}{2} \log(x^2 + a^2)$, i.e.
$f(x)= - U'(x)=-  \frac{\alpha x}{x^2 + a^2}$, and define the dimensionless parameters
$r=\frac{\alpha}{2 v_0 a}$, and $g=2 \gamma \alpha/v_0^2$ (see Fig. \ref{Fig_log}). We choose the period $[-L/2,L/2]$.
Note that there is then a jump of the force at $x= \pm L/2$ in order to ensure periodicity.
Since we are interested in the large $L$ limit, this jump is very small and does not change
our results below in that limit. 
One can calculate, denoting $\tilde x=x/a$ and $\tilde x_0 = x_0/a$, 
\be
W(x) - W(x_0) = - 2 \gamma \int_{x_0}^x dy \frac{f(y)}{v_0^2-f^2(y)} = \frac{g}{2}  \int_{\tilde x_0^2}^{\tilde x^2} \frac{dz (1+z)}{1 + 2 z(1- 2 r^2) + z^2} \;,
\ee 
which holds for (i) $r<1$ and any $x,x_0$ (ii) $r>1$ and $-y_s < x,x_0 < y_s$ with $y_{s,u}=a(r \mp \sqrt{r^2-1})$ (see Fig. \ref{Fig_log}).
For $r<1$ one sets $z= 2 r^2-1 + u' 2 r \sqrt{1-r^2}$ and $x= 2 r^2 -1 + u 2 r \sqrt{1-r^2}$ and obtains
\bea
W(x) - W(x_0) = 
\begin{cases} 
&\frac{g}{2}  \int_{u_0}^{u} du' \frac{u' + \frac{r}{\sqrt{1-r^2}}}{1 + (u')^2} \quad , \quad \tilde x^2= 2 r^2 -1 + u 2 r \sqrt{1-r^2} \quad , \quad r<1 \\
&\\
&\frac{g}{2}  \int_{\tilde x_0^2}^{\tilde x^2} \frac{dz (1+z)}{(1-z)^2} \quad , \quad ~~ \quad , \quad r=1\\
&\\
&\frac{g}{2}  \int_{u_0}^{u} du' \frac{u' + \frac{r}{\sqrt{r^2-1}}}{(u')^2-1} \quad , \quad \tilde x^2= 2 r^2 -1 + u 2 r \sqrt{r^2-1} \quad , \quad r>1 \end{cases}
\eea
where in the last formula $u,u_0<-1$ ($u=-1$ corresponds to $x=\pm y_s$). Performing the
integrals this leads to
\bea \label{rs1} 
\Phi_{\pm}(0,x)=  \frac{v_0^2}{v_0^2 - f^2(x)} e^{\mp  (W(x)-W(0))} =
\begin{cases}   
&\frac{(u + \frac{r}{\sqrt{1-r^2}})^2}{1 + u^2}  e^{\mp \frac{g}{2}( \frac{1}{2} \log(1+u^2) + \frac{r}{\sqrt{1-r^2}} \arctan u) \pm K } \quad , \quad  r<1  \\
& \\
& \frac{(u + \frac{r}{\sqrt{r^2-1}})^2}{u^2-1} 
 e^{\mp \frac{g}{2}( \frac{1}{2} \log(u^2-1) - \frac{r}{\sqrt{r^2-1}} {\rm arccoth} u) \pm K} \quad , \quad r>1 
\end{cases}
\eea
where $K$ is a $r$ dependent integration constant which cancels out in the observables of interest. Note that the ratio between the two species is
\be
\frac{\tilde P_-(x)}{\tilde P_+(x)} = \frac{v_0 + f(x)}{v_0 - f(x)} = \frac{x^2 + a^2 - 2 r x a}{x^2 + a^2 + 2 r x a} \;.
\ee 
For $r \geq 1$ it vanishes near the stable point $y_s$ for the $+$ species (where they outnumber
the $-$ species). \\

\noindent
{\bf Phase $A$}.
For $r<1$ (phase $A$) we thus obtain the stationary distribution as
\bea \label{ppm} 
\tilde P(x) = \tilde A \Phi_+(0,x) \quad , \quad 
\Phi_\pm(0,x)= \frac{(u + \frac{r}{\sqrt{1-r^2}})^2 }{(1+u^2)^{1 \pm \frac{g}{4}}}  e^{\mp \frac{g}{2} \frac{r}{\sqrt{1-r^2}} \arctan u \pm K} \quad , \quad u=\frac{(x/a)^2 + 1 - 2 r^2}{2 r \sqrt{1-r^2}}
\eea
where $K$ is an immaterial constant and $\tilde A e^K$ is determined by the normalisation condition
$\int_{-\infty}^{+\infty} dx \tilde P(x)=1$ \cite{footnote3}. Since $\Phi_+(0,x)
\propto |x|^{-g}$ for $|x| \to +\infty$, this stationary measure is normalisable
only for $g>1$. Hence for $g>1$ the particle is bound by the logarithmic potential, which corresponds to the 
phase bound $A$ in Fig. 4 of the text. In that phase $D_L \sim d_r L^{1-g}$ at large $L$ from \eqref{DL0}
and \eqref{ppm}. 
For $g<1$, the above stationary distribution is not
normalisable in the infinite space. This corresponds to the 
phase unbound $A$ in Fig. 4 of the text. Using $L$ as an upper cutoff one finds
from \eqref{DL0}, $D = D_0 (1-g^2)$ in that phase. 

\noindent{\bf Critical point}.
At $r=1$, i.e. the critical point between the $A$ and $C$ phases, the stationary distribution reads
for any $g>1$
\be \label{r1} 
\tilde P(x) = \tilde A \Phi_+(0,x) \quad , \quad \Phi_\pm(0,s)= \frac{(a^2 + x^2)^2}{(a^2-x^2)^{2 \pm \frac{g}{2}}} e^{\mp \frac{g a^2}{a^2 - x^2} \pm K}
\quad , \quad -a < x < a
\ee
and vanishes outside the interval $[-a,a]$. It thus vanishes with an essential singularity at $x=\pm a$.
This distribution is bimodal for $g<8$ and, in that case, with two symmetric maxima close to the boundaries at $x = \pm a$, see Fig. \ref{Fig_r1}. 
\begin{figure}
\includegraphics[width = 0.9\linewidth]{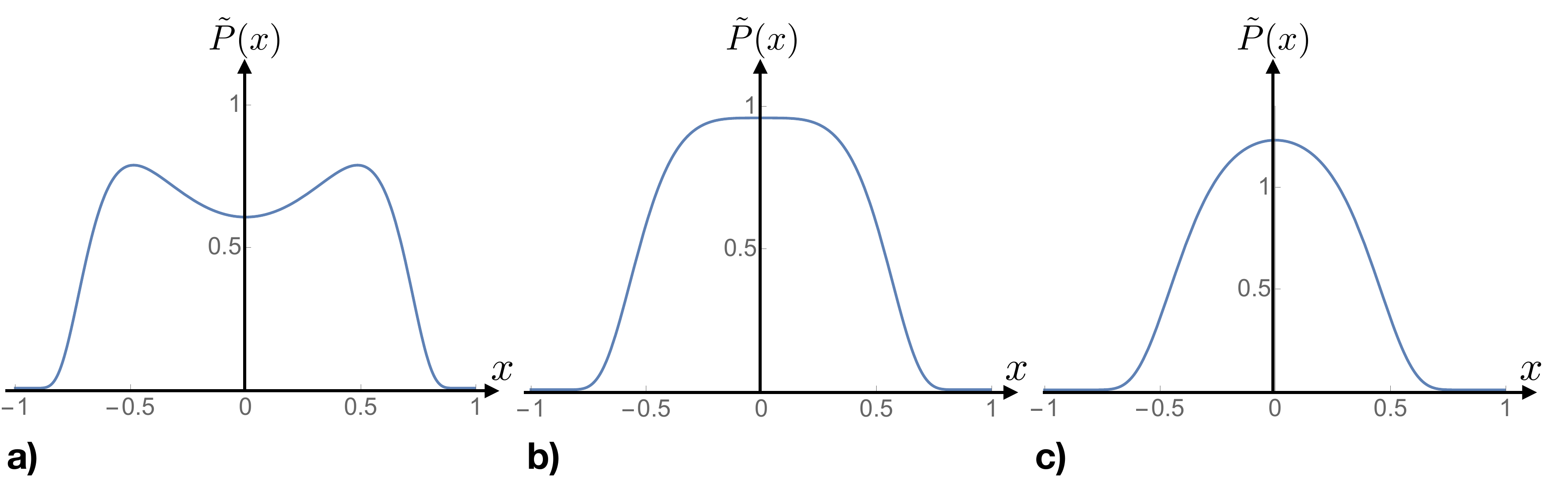}
\caption{Plot of the stationary distribution $\tilde P(x)$ corresponding to the force $f(x) = - \alpha x/(x^2+a^2)$ for $a=1$ (see Fig. \ref{Fig_log}) at the critical point $r=1$, as given by the exact formula in Eq. (\ref{r1}). It exhibits a shape transition as $g$ crosses the value $g_c=8$. For $g=4<g_c$, $\tilde P(x)$ is bi-modal (panel {\bf a}) with two maxima close to the boundaries at $x = \pm 1$ while for $g=12>g_c$, $\tilde P(x)$ is unimodal with a maximum at $x=0$ (panel {\bf c}). Exactly at the critical value $g=g_c=8$ (panel {\bf b}), the stationary distribution has a very flat maximum at $x=0$.}\label{Fig_r1}
\end{figure}

\noindent{\bf Phase $C$}. For $r>1$ the formula \eqref{rs1} gives, up to a normalisation prefactor,
the stationary distribution $\tilde P(x) = \tilde A \Phi_+(0,x)$ in the bound phase $C$
(see Fig. 4 in the text).
The stationary distribution has a finite support, which is the interval $[x_s, y_s]$ 
with $y_s=a(r - \sqrt{r^2-1})>0$ and $x_s=-y_s$. It is an even function of $x$ and vanishes at the edges of
the support (stable fixed points) as a power law, $\tilde P(x) \sim (y_s-x)^\phi$. Note that for $r>1$, $f'(y_s)= - \frac{2 r v_0}{a}  \frac{r \left(\sqrt{r^2-1}+r\right)-1}{2 r^2}$, 
hence the singularity exponent $\phi$, given in \eqref{phi} in the main text, reads
\be
\phi = - 1 + \frac{g}{4} \frac1{r \left(\sqrt{r^2-1}+r\right)-1} = - 1 + \frac{g}{4} \left(\frac{r}{\sqrt{r^2-1} } -1\right) 
\ee 
Within the phase $C$ there is thus a transition line $g=g_c(r)=4 \left(r \left(\sqrt{r^2-1}+r\right)-1\right)$
where the exponent $\phi$ changes sign (for $g<g_c(r)$ $\phi<0$).\\


\noindent
{\bf Vanishing of diffusion constant at the transition from phase $A$ to phase $C$}.
Note that in the bound phase $C$ the diffusion constant vanishes
$D_L=0$ for  any $L > 2 y_s$ since the particle is blocked
by the first absorbing region that it encounters (remembering that the
force $f(x)$ is chosen periodic, hence the RTP will move at most by one period). 
Let us see how it vanishes as $r \to 1^-$, coming from the bound $A$ phase. 
From \eqref{DL0} we see
that we can take the constants $K=0$ since they cancel in the product. Let us
consider $g>1$, i.e. the bound $A$ phase. From \eqref{rs1} for $r<1$ we see that $u \simeq \frac{\tilde x^2-1}{2 \sqrt{2(1-r)}}$ as $r \to 1^-$,
hence for any fixed $\tilde x \neq 1$ we can use the large $u \to \pm \infty$ asymptotics. Hence we have, as $r \to 1^-$
\be \label{1}
\Phi_\pm(0,x) \simeq  (2 \sqrt{2(1-r)})^{\pm g/2} 
\frac{(\tilde x^2 + 1)^2}{(\tilde x^2-1)^{2\pm\frac{g}{2}}} 
e^{\mp  {\rm sgn}(\tilde x^2-1) \frac{\pi g}{4} \frac{1}{\sqrt{2 (1-r)} } }  e^{\pm \frac{g}{\tilde x^2-1}} \;.
\ee
From \eqref{DL0}, in order to calculate $D_L$ we need to compute the integrals
$2 \int_0^{L/2} dx  \, \Phi_\pm(0,x)$ (recalling that the functions $\Phi_\pm(0,x)$ are even). Both integrals
are found to be exponentially diverging as $\sim e^{\frac{\pi g}{4} \frac{1}{\sqrt{2 (1-r)} } }$,
for $\Phi_+(0,x)$ from the region $\tilde x<1$ and for 
$\Phi_-(0,x)$ from the region $\tilde x>1$. Indeed one has
\bea
&& 2 \int_0^{L/2} dx  \Phi_\pm(0,x) \simeq 2 \, e^{\frac{\pi g}{4} \frac{1}{\sqrt{2 (1-r)} } } 
a  \left(2 \sqrt{2(1-r)}\right)^{\pm g/2} 
\int_{\tilde x \in I_\pm} d\tilde x \frac{(\tilde x^2 + 1)^2}{(\tilde x^2-1)^{2\pm\frac{g}{2}}} 
e^{\pm \frac{g}{\tilde x^2-1}} 
\eea
where $I_+=[0,1]$ and $I_-=[1,L/a[$. Hence we obtain
\be
\frac{D_0}{D_L} \simeq 4 a^2 L^{-2} e^{2 \frac{\pi g}{4} \frac{1}{\sqrt{2 (1-r)} } } 
\int_{0}^1 d\tilde x \frac{(\tilde x^2 + 1)^2}{(\tilde x^2-1)^{2+\frac{g}{2}}} 
e^{- \frac{g}{1-\tilde x^2}} 
\times \int_{1}^{L/a} d\tilde x \frac{(\tilde x^2 + 1)^2}{(\tilde x^2-1)^{2-\frac{g}{2}}} 
e^{- \frac{g}{\tilde x^2-1}} \sim \left(\frac{L}{a}\right)^{g-1} e^{-\frac{\pi g}{2} \frac{1}{\sqrt{2 (1-r)} } } \;. \label{essential}
\ee
We see that the first integral is convergent, while the second behaves
at large $L/a$ as $(L/a)^{1+g}$. Putting all factors together we obtain that 
$\frac{D_L}{D_0} \sim (L/a)^{1-g} e^{-\frac{\pi g}{2} \frac{1}{\sqrt{2 (1-r)} } }$
with an amplitude that can be obtained from the above integrals. 
This leads, for $g>1$, to the essential singularity in the limit $r \to 1^-$ for the diffusion constant, as 
given in the text. Note that it
is multiplied by the estimate for the diffusion constant in the unbound $A$ phase, $(L/a)^{1-g}$. \\

Let us indicate a qualitative, but more general argument for the vanishing of the diffusion constant upon entering phase $C$, valid for
any smooth enough $f(x)$ (twice differentiable). For simplicity and by choice of coordinate (thus without loss of generality), we fix
the position of the minimum of $f(x)$ to be at $x=0$, and define $f(0)=- v_0(1-\rho)$.
Upon approaching phase $C$, coming from phase $A$, we have $\rho \to 0^+$.
In that limit, we can approximate $f(y)=- v_0(1-\rho) + \frac{1}{2} f''(0) y^2$, with $f''(0)>0$.
One has, denoting $y=\rho z \sqrt{v_0/f''(0)}$, the following divergence of the active potential
\bea \label{WW} 
W(x) - W(x_0) \simeq
2 \gamma  \int dy \frac{1}{2 \rho v_0 + f''(0) y^2} \simeq 
\frac{2 \gamma}{\sqrt{2 \rho v_0 f''(0)}} \int dz \frac{1}{1+ z^2} 
\eea 
where for any fixed $x, x_0$, the boundaries of the integrals are pushed to infinity. If one integrates the last integral in \eqref{WW} 
from $- \infty$ to $+\infty$ (which corresponds to $x, x_0$ each on one side of $x=0$) and substitute in the formula for the diffusion constant, one finds again the estimate
$D_L \sim e^{-  \frac{\pi g}{2} \frac{1}{\sqrt{2 (1-r)} } }$, in agreement with the previous computation~(\ref{essential}).

%
%
%
%
%


\medskip
\begin{center}
{\bf E. Mean first-passage time}
\end{center}
\medskip

We consider the particle in $x \in [-L_1,+\infty)$ and denote $T_{\pm}(x)$ the mean first-passage times to the level $X$, starting initially at $x$ in the velocity state $\sigma=\pm$. We focus on phase $A$ here, i.e. 
$|f(x)|<v_0$. Note that here we do not assume any periodicity of $f(x)$. For convenience, we consider reflecting
boundary conditions at $x=-L_1$. 
Then $T_\pm(x)$ satisfy the following pair of backward Fokker-Planck equations in $x\in [-L_1, X]$
\bea
\left[v_0+ f(x)\right]\, \frac{dT_+}{dx}- \gamma\, T_+(x)+ \gamma\, T_{-}(x) &= & -1 \label{Tplus.1} \;, \\
\left[-v_0+ f(x)\right]\, \frac{dT_-}{dx}- \gamma\, T_-(x)+ \gamma\, T_{+}(x) &= & -1 \label{Tminus.1} \;.
\eea
The boundary conditions for $T_{\pm}(x)$ are 
\be \label{bcL1} 
T_+(X)=0 \quad , \quad \frac{d T_-(x)}{dx}\Bigg |_{x=-L_1}=0  \;.
\ee 
The first condition comes from the fact that if the particle starts at $X$ with a positive state $\sigma=+1$,
and given that $|f(L)|<v_0$, it implies that the particle crosses $X$ immediately. The second condition is more tricky to derive. By writing down the backward Fokker-Planck equation exactly at $x=-L_1$ and imposing that if the particle tries to go to the left of $-L_1$ it remains stuck at $x=-L_1$, we can show that this second condition emerges. Since eventually we will be interested in the limit $L_1 \to - \infty$, with a non negative drift, this second boundary condition is expected to be unimportant.

One can eliminate $T_{+}(x)$ and write a single differential equation for $T_-(x)$ as follows. We re-write the two equations
in the operator form:
\bea
{\cal L}_+ T_+(x) \equiv \left[(v_0+f(x)) \partial_x-\gamma\right] T_+(x)= -1-\gamma\, T_{-}(x) \;, \label{Tplus.2} \\
{\cal L}_- T_-(x) \equiv \left[(-v_0+f(x)) \partial_x-\gamma\right] T_-(x)= -1-\gamma\, T_{+}(x) \, .  \label{Tminus.2}
\eea
Operating with ${\cal L}_{+}$ on the left and right hand side of Eq. (\ref{Tminus.2}) 
and using \eqref{Tplus.2} we get, denoting $Z_-(x)= dT_-/dx$,
the first order differential equation 
\be
\left[v_0^2-f^2(x)\right]\, \frac{dZ_-}{dx} +  \left[2\,\gamma\, f(x)- v_0\, f'(x) - f(x)\, f'(x)\right]\, Z_-(x)=-2\gamma\, .
\label{Wx.1}
\ee
%
Integrating, and taking into account the boundary condition at $x=-L_1$ \eqref{bcL1}, we obtain
\be
Z_-(x)= \frac{dT_-}{dx}= - \frac{2\gamma}{v_0-f(x)}\, \int_{-L_1}^x \frac{dy}{v_0+f(y)}\, 
e^{-2\gamma\, \int_y^x \frac{f(u)\, du}{v_0^2-f^2(u)} } \;.
\label{Wx.2}
\ee
Integrating we get
\be
T_-(x) = \int_{-L_1}^x dy \, Z_-(y) + c 
\ee 
where $c$ is an arbitrary constant yet to be fixed. Substituting this result into Eq. \eqref{Tminus.1}
we obtain $T_+(x)$. Using the boundary condition $T_+(X)=0$ gives us the
constant $c$, and finally we obtain
\be
T_-(x) = T_-(X) + \int_x^X dy \frac{2 \gamma}{v_0-f(y)} \int_{-L_1}^y \frac{dz}{v_0 + f(z)} e^{-2\gamma\, \int_z^y \frac{f(u)\, du}{v_0^2-f^2(u)} } \;,
\ee 
where 
\be
T_-(X) = \frac{1}{\gamma} \left( 1+ 2 \gamma
\int_{-L_1}^L \frac{dy}{v_0 + f(y)} e^{-2\gamma\, \int_y^L \frac{f(u)\, du}{v_0^2-f^2(u)} } \right) \;.
\ee 
Interestingly, note that $T_-(X)$, which is the mean first return time to $x=X$ of a RTP
starting in the state $\sigma=-1$ (which implies that it travels first to the left),
is non zero, contrarily to the diffusive limit where both $T_\pm(X)$ vanish.
From this result we also obtain
\be
T_{-}(x) - T_{+}(x) = 2  \int^x_{-L_1} \frac{dy e^{W(x)-W(y)}}{v_0+f(y)} + \frac{1}{\gamma} \;.
\ee
In the case where the bias is to the right (i.e. meaning here $W(-\infty) \to + \infty$ where $W(x)$
is defined in Eq. \eqref{Wdef} of the main text), we can safely take $- L_1 \to - \infty$ and obtain our final result for the first mean passage time from $x$ to $X$ on the infinite line in phase $A$, as given in the text [see Eqs. (\ref{Tplus_final}) as well as (\ref{deltaTpm})].

We now show that if there is a finite velocity $V$ in the large $X$ limit, which can be extracted from
the mean first passage time as
\be
\lim_{X \to +\infty} \frac{T_-(x)}{X} = \frac{1}{V} 
\ee 
for arbitrary fixed $x$ then it coincides with the one obtained in formula \eqref{VelInfty} of the text.

To show this we choose $x=0$, and we can also choose $L_1=0$, as it does not
affect the (positive) velocity in the large $X$ limit. We obtain
\be
\frac{1}{X} T_-(0) \simeq 
\frac{1}{X}  \int_0^X dy \frac{2 \gamma}{v_0-f(y)} \int_{0}^y \frac{dz}{v_0 + f(z)} e^{-2\gamma\, \int_z^y \frac{f(u)\, du}{v_0^2-f^2(u)} }
\ee 
since $T_-(X)$ remains bounded as $X \to +\infty$. 
Multiplying the numerator and the denominator by $(v_0+ f(y))(v_0-f(z))$ we obtain
\be
\frac{1}{X} T_-(0) \simeq 
\frac{1}{X}  \int_0^X dy \frac{2 \gamma}{v^2_0-f^2(y)} \int_{0}^y \frac{dz}{v_0^2 - f(z)^2} 
(v_0^2 + f(y) v_0 - f(z) v_0 - f(y) f(z)) 
e^{-2\gamma\, \int_z^y \frac{f(u)\, du}{v_0^2-f^2(u)} } \;,
\ee
which we can then expand into the sum of four terms. One can check that the first term is
exactly the first term in~\eqref{VelInfty}. We can further show using integration by parts that the sum of the second and 
the third term gives the second term in \eqref{VelInfty}. Finally the fourth term vanishes in
the large $X$ limit, using again integration by parts. This shows that the exchange of limits
(large spatial period and large time) used to
obtain \eqref{VelInfty} is legitimate. 

\medskip
\begin{center}
{\bf F. Fully inhomogeneous model}
\end{center}
\medskip

All our formula extend easily to the case where the RTP velocity $v_0$ and transition rate
$\gamma$ depend also on space $v_0 \to v_0(x)$ and $\gamma \to \gamma(x)$, with the
same periodicity of period $L$.
The model is now defined by the pair of Fokker-Planck equations which generalise the 
Eqs. \eqref{P-evol.1}
\bea
&& \partial_t P_+ = - \partial_x [ (f(x) + v_0(x))  P_+]  -  \gamma(x) P_+ + \gamma(x) P_-  \label{P+evol.1}\\
&& \partial_t P_- = - \partial_x [ (f(x) - v_0(x))  P_- ] + \gamma(x) P_+ -  \gamma(x) P_-  \, . 
\label{P-evol.12}
\eea
We can now follow all the steps presented in the paper and check that they generalise straightforwardly to this fully inhomogeneous case. 
Let us illustrate it in the case of phase $A$, $|f(x)| < v_0(x)$ (the phases $A$, $B$, $C$, $D$ generalise also straightforwardly
and we assume that $v_0(x)>0$, $\gamma(x)>0$ for all $x$). The equations \eqref{tildeP} still hold with the substitutions $v_0 \to v_0(x)$ and $\gamma \to \gamma(x)$. From them one obtains that the
equations for the stationary distributions now read
\bea \label{stateq2} 
&& f(x) \tilde P(x) + v_0(x) \tilde Q(x) = J \\
&& \frac{d}{dx} [\frac{v^2_0(x) - f^2(x)}{v_0(x)} \tilde P(x) + J \frac{f(x)}{v_0(x)}] 
+ 2 \frac{\gamma(x)}{v_0(x)}  J - 2 \frac{\gamma(x) f(x)}{v_0(x)} \tilde P(x)  = 0 
\eea
where the second equation generalises Eq. \eqref{eq3}. Solving it one finds that for $J \neq 0$, the formula \eqref{P0} of the text for the stationary distribution becomes
\be \label{P02} 
 \tilde P(x) = \frac{J}{v^2_0(x) - f^2(x)} \left(\int_0^L \frac{du \, 2 \,\gamma(u) \Phi_-(x,u)}{A_L}
- \int_0^x du \, 2 \, \gamma(u) \Phi_-(x,u) - f(x) \right) 
\ee
with now $\Phi_{\pm}(x,u)=  \frac{v_0(x) v_0(u)}{v_0(u)^2 - f(u)^2} e^{\pm  (W(x)-W(u))}$
and $A_L$ given by \eqref{AL}, 
where from now on the definition
of the ``active potential'' \eqref{Wdef} becomes
\be \label{Wdef2} 
W(x)  = - 2 \int_0^x dy \frac{\gamma(y) f(y)}{v^2_0(y)-f^2(y)} \;.
\ee 
The formula for the velocity $V_L$ remains the same as \eqref{VL} 
where one substitutes in the last term $\frac{W(L)}{2 \gamma L}  \to 
\frac{1}{L} \int_0^L dx \frac{f(x)}{v^2_0(x) - f^2(x)}$ and 
the function $\Psi(x,u)$ is now 
$\Psi(x,u)=  \frac{2 \gamma(u) v_0(x) v_0(u) \, e^{- (W(x)-W(u))} }{(v^2_0(x) - f^2(x))(v_0^2(u) - f^2(u))}$.\\

Consider now the diffusion constant in the absence of a bias, i.e. $W(0)=W(L)$. The equations 
\eqref{P+evol.3}, \eqref{P-evol.2}, \eqref{RS}, \eqref{rs0} again still hold with the substitution $v_0 \to v_0(x)$ and $\gamma \to \gamma(x)$. From \eqref{stateq2}, the equation for the stationary distribution in the case $J=0$ is slightly modified as compared to \eqref{At} and becomes
\bea
\tilde P(x) =  \tilde A \frac{v_0(x)}{v^2_0(x) - f^2(x)} 
e^{2  \int_0^x dy \frac{\gamma(y) f(y)}{v^2_0(y) - f^2(y)}} \quad , \quad 
\frac{1}{\tilde A} = \int_0^L dx \frac{v_0(x)}{v^2_0(x) - f^2(x)} 
e^{2  \int_0^x dy \frac{\gamma(y) f(y)}{v_0^2(y) - f^2(y)}}  \label{Att}
\eea 
The equations \eqref{rs2}, \eqref{rs20} and \eqref{rrr} still hold setting $v_0 \to v_0(x)$ and $\gamma \to \gamma(x)$.
The equation \eqref{eqq} becomes
\be \label{eqq2} 
0 = - \frac{d}{dx} \left[ \frac{f^2(x) - v^2_0(x)}{f(x)} r(x) + c \frac{v_0(x)}{f(x)} \right] - 2 \gamma(x) r(x)
+ \tilde A \,
e^{2  \int_0^x dy \frac{\gamma(y) f(y)}{v_0^2(y) - f^2(y)}} \;.
\ee
Defining again \eqref{rtilde}, 
integrating and using the periodicity to determine the integration constant $c$, we obtain
after an integration by part, the following expression for the diffusion constant $D_L$ in the fully inhomogeneous case at zero bias
\bea \label{DLSM2} 
D_L = \frac{L^2}{\left( \int_0^L dx \frac{2 \gamma(x) v_0(x)}{v^2_0(x)- f^2(x)}  
e^{- 2  \int_0^x dy \frac{\gamma(y) f(y)}{v_0^2 - f^2(y)}} \right) 
\left( \int_0^L dx \frac{v_0(x)}{v^2_0(x)- f^2(x)}  
e^{2  \int_0^x dy \frac{\gamma(y) f(y)}{v_0^2 - f^2(y)}} \right)}
\eea 
Note the non-trivial limit for the diffusion constant in the absence of an external force $f(x)=0$ but in the presence of inhomogeneities in the
velocity and rate
\be
D_L = \frac{L^2 }{\left( \int_0^L dx \frac{2 \gamma(x)}{v_0(x)} \right) 
\left( \int_0^L dx \frac{1}{v_0(x)}   \right)} \;,
\ee 
as given in Eq. (\ref{Dinh}) in the text. 

For the mean first passage time, Eqs. \eqref{Tplus.1}, \eqref{Tminus.1}, 
\eqref{Tplus.2}, \eqref{Tminus.2},
still hold with the substitutions $v_0 \to v_0(x)$ and $\gamma \to \gamma(x)$. We need now to operate with
${\cal L}_{+}$ on the left and right hand side of Eq. (\ref{Tminus.2}) 
divided by $\gamma(x)$. This leads to the following equation for $Z_-(x)= dT_-/dx$, 
which generalises \eqref{Wx.1}
\bea
&& \left[v^2_0(x)-f^2(x)\right]\, \frac{dZ_-}{dx} +  \left[2\,\gamma(x) \, f(x)
- \gamma(x) (v_0(x)+f(x)) \frac{d}{dx} \left[ \frac{- v_0(x)+f(x)}{\gamma(x)}\right] \right] 
 Z_-(x) \\
 && =-2\gamma(x) + \gamma(x) (v_0(x) + f(x)) \frac{d}{dx} \left[ \frac{1}{\gamma(x)} \right]
\label{Wx.1new}
\eea
This leads to 
\be
Z_-(x)= \frac{dT_-}{dx}= - \frac{2\gamma(x)}{v_0(x)-f(x)}\, \int_{-L_1}^x dy \left(\frac{1}{v_0(y)+f(y)} - \partial_y \frac{1}{2 \gamma(y)} \right)
e^{-2\, \int_y^x \frac{\gamma(u) f(u)\, du}{v_0^2(u)-f^2(u)} }
\label{Wx.2new}
\ee
Hence we obtain
\be\label{Tminh}
T_-(x) = T_-(L) + \int_x^L dy \frac{2 \gamma(y)}{v_0(y)-f(y)} \int_{-L_1}^y dz \left( \frac{1}{v_0(z) + f(z)} 
- \partial_z \frac{1}{2 \gamma(z)}  \right) e^{-2\, \int_z^y \frac{\gamma(u) f(u)\, du}{v_0^2(u)-f^2(u)} } \;.
\ee 
Substituting this form (\ref{Tminh}) in Eq. (\ref{Tminus.2}) we obtain $T_+(x)$ and by imposing the boundary condition $T_+(L) = 0$, we finally obtain 
\be
T_-(L) = \frac{1}{\gamma(L)} + 2 
\int_{-L_1}^L dy \left(\frac{1}{v_0(y)+f(y)} - \partial_y \frac{1}{2 \gamma(y)} \right)
e^{-2\, \int_y^L \frac{\gamma(u) f(u)\, du}{v_0^2(u)-f^2(u)} } \;.
\ee 
%
In presence of a bias to the right one can take the limit $L_1 \to -\infty$,
which generalises the formula in the text.

\end{widetext}


\begin{thebibliography}{}
\bibitem{footnote11} For uniform $f(x)=f$ it gives $V=f$, as
obvious from~\eqref{model}.
\bibitem{footnote2} $f'(x)$ is 
discontinuous but the integral \eqref{P2} 
is well defined.
\bibitem{footnote3} For $L/a \gg 1$ we can neglect the
small jump in $f(x)$ at~$x~=~\pm L/2$.
\end{thebibliography}
\end{document}